\documentclass[11pt]{JHEP3}
\let\ifpdf\relax
\usepackage{ifpdf}

\JHEPspecialurl{http://jhep.sissa.it/JOURNAL/JHEP3.tar.gz}

\usepackage{epsfig,multicol,amsmath}

\usepackage{epstopdf}
\usepackage{bm}  
\usepackage{amsmath}     
\usepackage{epsfig,epsf}
\usepackage{rotating}

\usepackage{cite}

\newcommand{\bas}{\bar{\alpha}_S}
\newcommand{\as}{\alpha_S}

\newcommand{\Lb}{\left(}
\newcommand{\Rb}{\right)}
\parskip=\itemsep               

\newcommand{\nn}{\nonumber}
\newcommand{\D}{\partial}

\newcommand{\ga}{\gamma}

\newcommand{\Ga}{\Gamma}
\newcommand{\om}{\omega}

\begin{boldmath}
\title{Semiclassical solution to the BFKL equation with massive gluons}
\end{boldmath}
\author{\Large 
Eugene Levin,${}^{a,b}$ \,Lev  Lipatov${}^{c}$\, and\, Marat Siddikov${}^{b}$\\
 ${}^a$\, 
 Department of Particle Physics, School of Physics and Astronomy,
Tel Aviv University, Tel Aviv, 69978, Israel\\
${}^b$ \, Departamento de F\'\i sica,
Universidad T$\acute{e}$cnica Federico Santa Mar\'\i a   and
Centro Cient\'\i fico-Tecnol$\acute{o}$gico de Valpara\'\i so,
Casilla 110-V,  Valparaiso, Chile\\
${}^c$\,Theoretical Physics Department,
Petersburg Nuclear Physics Institute,
Orlova Roscha, Gatchina,
188300, St. Petersburg, Russia}

\abstract
{In this paper we proceed  to study the high energy behavior of  scattering amplitudes in a  simple  field model, with the Higgs mechanism for the gauge boson mass.  The spectrum of the $j$-plane singularities of the $t$-channel partial waves,  and corresponding eigenfunctions of the BFKL equation in leading log($1/x$) approximation (LLA),  were previously calculated  numerically.
Here a semiclassical approach is developed to investigate the influence of the impact parameter exponential decrease, existing in this model, on the high energy asymptotic behaviour of the scattering amplitude. This approach is much simpler than the numerical calculations, and  reproduces their results.
The  analytical (semi-analytical) solutions which have been found in the approximation, can be used to   incorporate  correctly the  large impact parameter behavior in the framework of CGC/saturation approach.
This behaviour is  interesting as  provides the high energy amplitude for the electroweak theory, which can be measured experimentally.}
\dedicated{PACS: 12.38-t, 12.38.Cy,1 2.38.Lg, 13.60.Hd, 24.85.+p, 25.30.Hm}
\preprint{TAUP  \\
{\tt USM-TH-337}\\
\today}

\begin{document}

\section{ Introduction}
In our paper~\cite{LLS} we solved  the BFKL equation with  a massive gluon in the framework of the Higgs model numerically.
Such an  equation arises in the electroweak theory with zero  Weinberg angle   ( see Ref. ~\cite{BLP}). However,  for us, this gauge invariant theory  provides an instructive example of a model  in which the scattering amplitude has correct large impact parameter behavior ( scattering amplitude $\propto \exp\left( - m\, b \right)$ at large $b$), but we still have  the unitarity problem as the scattering amplitude increases as $s^\Delta$ at high energies. Therefore, this model  is a perfect training ground to study how the correct $b$ behavior,  can influence the resolution of the unitarity problem in  the framework of the CGC/saturation approach ~\cite{GLR,MUQI,MV,REV}. 
This approach leads to a  partial amplitude smaller than unity, as required by unitarity constraints. However, it generates a radius of interaction that increases as
power of energy ~\cite{KW1,KW2,KW3,FIIM}, 
leading to the violation of the Froissart bound~\cite{FROI}.

Another facet of this model, is that it is a  possible candidate for an  effective theory equivalent to perturbative  QCD,  in the region of distances ($r$) shorter than $1/m$, where $m$ denotes the gluon mass. Indeed, for $r \,\ll\,1/m$, the fact that gluon has a mass, is not essential  while at $r \sim 1/m$ similar correlation functions
 arise from fixing or eliminating Gribov's copies ~\cite{GRCO} 
 (see Refs. ~\cite{GRCTH,GRCREV,GRCMASSG}). We wish to stress that a  gauge theory with the  Higgs mechanism, leads to a  good description of the gluon propagator, 
 calculated in lattice approach~\cite{GRCLAT} with  $m = 0.54 \,GeV$. Therefore, 
 a plausible scenario is that 
 the Higgs gauge theory  describes QCD  in the kinematic region $r\,\leq 1/m$, while for $r \sim 1/\Lambda_{\mbox{\tiny QCD}} \,>\,1/m$ the non-perturbative QCD approach takes over,  and leads to the confinement of quarks and gluons, which is missing in the theory with a massive gluon.

We found~\cite{LLS} that the spectrum of the massive BFKL equation in $\omega$-space for $t=0$,  coincides with that of  massless one~~\cite{BFKL,LIREV}. The  simple parametrization  of the eigenfunctions have also been obtained in Ref.~\cite{LLS}. In this paper we propose using the solution to the BFKL equation with massive gluons which has an advantage of  being simple and semi-analytic.  Having  this solution in hand, we are able to progress to more difficult problems, e.g. a generalization of the main equations of the CGC/saturation approach~\cite{BK,JIMWLK,REV}.

\begin{figure}[ht]
     \begin{center}
     \includegraphics[width=12cm]{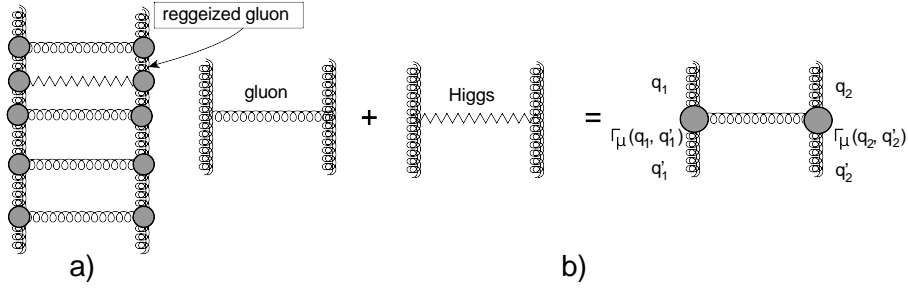} 
     \end{center}    
      \caption{ The massive BFKL equation (\protect Fig.~\protect\ref{eq}-a) and its kernel (\protect Fig.~\protect\ref{eq}-b)  }
\label{eq}
   \end{figure}

 The equation for the scattering amplitude in leading $\ln(1/x)$ approximation of perturbative theory, has been on the market for some time~~\cite{BFKL}, and is schematically shown in the Fig.~\protect\ref{eq}. Its kernel is of the  form~~\cite{BFKL,LLS}
 
    \begin{eqnarray} \label{K0}
    K\left( q_1, q_2 | q'_1, q'_2\right)\,\,
    &=& \frac{\as N_c}{2 \pi^2}\Big\{ \frac{1}{k^2 + m^2}\Big(\frac{q^2_1 + m^2}{q'^2_1 + m^2}\,+\,\frac{q^2_2 + m^2}{q'^2_2 + m^2}\Big)\,\,-\,\,\frac{q^2 +\frac{N^2_c+1}{N^2_c} m^2}{(q'^2_1 + m^2) ( q'^2_2 + m^2)}\Big\}.
    \end{eqnarray}   
    For $q=0$, the kernel~(\ref{K0}) simplifies considerably,  and yields a homogeneous BFKL equation for the Yang-Mills theory with the Higgs  mechanism,
    \begin{equation} \label{EQ}
\om f\left( p\right)\,\,=\,\,2 \om\left( p \right) f\left( p\right)\,+\, \frac{\as N_c}{2 \pi^2}\int d^2 p' \Big( \frac{ 2 f\left( p'\right)}{\left( \vec{p}  - \vec{p}^{\,\,'}\right)^2  +  m^2}\,\,-\,\, \frac{\frac{N^2_c+1}{N^2_c} m^2\,f\left( p'\right)}{(p^2 +m^2) (p'^2 + m^2}\Big),
\end{equation}
where $ q_1 = q_2 = p$, $q'_1 = q'_2 = p'$ and $\omega(p)$ is the gluon Regge trajectory given by
\begin{eqnarray} \label{GTR}
\om\left( p \right) \,\,&=&\,\,- \frac{\as N_c}{4 \pi^2}\int \frac{ d^2 k \left( p^2 + m^2\right)}{\left( k^2 + m^2\right) \left(\left(  \vec{p} - \vec{k}\right)^2 + m^2 \right)}\,=\, -\, \frac{\as N_c}{2 \pi^2}\frac{p^2 + m^2}{|p|\sqrt{p^2 + 4  m^2}}\ln \frac{\sqrt{p^2 + 4 m^2} + |p|}{\sqrt{p^2 + 4 m^2} - |p|}.
\end{eqnarray}
Examining the rotationally symmetric solution, the kernel can
be integrated over the azimuthal angle $\phi$.
Introducing the new variables
\begin{equation} \label{VAR}
\kappa\,=\,\frac{p^2}{m^2};\,\,\,\,\,\,\,\,\,\,\,\,\,\,\kappa'\,=\,\frac{p'^2}{m^2};\,\,\,\,\,\,\,\,\,\,\,\,\,E\,=\,- \frac{\om}{\bas};\,\,\,\,\,\,\,\,\,\,\,\,\, \bas\,=\,\frac{\as N_c}{\pi},
\end{equation}
and changing the  notation of the wave function $f(p)$ to $\phi_E(\kappa)$, we obtain the one-dimensional BFKL equation
\begin{eqnarray} \label{EQF}
\hspace{-0.5cm}&&E \phi_E\left( \kappa\right)\,\,=\,\,T\left( \kappa\right)\,\phi_E\left( \kappa\right)\,\,-\,\,\int^{\infty}_{0}\,\frac{d \kappa' \,\phi_E\left( \kappa'\right)}{\sqrt{( \kappa - \kappa')^2\,+\,2 (\kappa + \kappa') + 1}}\,\,+\,\,\frac{N^2_c + 1}{2 N^2_c}\frac{1}{\kappa + 1}\int^{\infty}_0 \frac{\phi_E\left( \kappa'\right) \,d \kappa'}{\kappa' + 1},
\end{eqnarray}
where the kinetic energy is given by
\begin{equation} \label{T}
T\left( \kappa\right)\,\,=\,\,\frac{\kappa + 1}{\sqrt{\kappa}\sqrt{\kappa + 4  }}\ln \frac{\sqrt{\kappa + 4 } + \sqrt{\kappa}}{\sqrt{\kappa + 4 } - \sqrt{\kappa}}.
\end{equation}
In this paper we will  deal mostly  with the wave function in $Y$ representation defined by a Fourier transform,
\begin{equation} \label{WFY}
\Psi\left( Y, \kappa\right)\,=\,\int^{\epsilon + i \infty}_{\epsilon - i \infty}\frac{d E}{2 \pi i}
e^{-E Y}\,\phi_E\left( \kappa\right).
\end{equation}
 For $\Psi\left( Y, \kappa\right)$, the Eq.~(\ref{EQF}) takes the form
 \begin{eqnarray} \label{EQY}
 \hspace{-0.5cm}&&\frac{\partial  \Psi\left( Y, \kappa\right)}{\partial Y} = - T\left(\kappa\right) \Psi\left( Y, \kappa\right) + \int^{\infty}_{0}\,\frac{d \kappa'\Psi\left( Y, \kappa'\right)}{\sqrt{( \kappa - \kappa')^2\,+\,2 (\kappa + \kappa') + 1}} - \frac{N^2_c + 1}{2 N^2_c}\frac{1}{\kappa + 1}\int^{\infty}_0 \frac{ \Psi\left( Y, \kappa\right)\,d \kappa'}{\kappa' + 1}
 \end{eqnarray}
  For completeness of presentation, we recall that the massless BFKL equations have the following form
 \begin{eqnarray} \label{EQBFKL}
E \,\phi^{\mbox{\tiny BFKL}}_E\left( \kappa\right)\,\,&\underset{\epsilon\,\to\,0}{=}&\,\,\ln \left( \frac{\kappa}{\epsilon}\right)\, \phi^{\mbox{\tiny BFKL}}_E\,\,-\,\,\int^{\infty}_{0}\,\frac{d \kappa' \,\phi^{\mbox{\tiny BFKL}}_E\left( \kappa'\right)}{|\kappa\,-\,\kappa'|\,+\,\sqrt{\kappa\,\epsilon}}=
-\,\,\int^{\infty}_{0}\,\frac{d \kappa' \,\left(\phi^{\mbox{\tiny BFKL}}_E\left( \kappa'\right) - \phi^{\mbox{\tiny BFKL}}_E\left( \kappa'\right)\right)}{|\kappa\,-\,\kappa'|},\nn\\
\frac{\partial \Psi^{\mbox{\tiny BFKL}}\left( Y ,\kappa\right)}{\partial  Y}\,\,&\underset{\epsilon\,\to\,0}{=}&\,\,-\ln \left(\frac{\kappa}{\epsilon}\right)\, \Psi^{\mbox{\tiny BFKL}}\left( Y ,\kappa\right)\,\,+\,\,\int^{\infty}_{0}\,\frac{d \kappa' \,\Psi^{\mbox{\tiny BFKL}}\left( Y, \kappa'\right)}{|\kappa - \kappa'|\,+\,\sqrt{\kappa\,\epsilon}}
\nn
\end{eqnarray}
The way of regularization at $\epsilon \to 0$ stems directly from Eq.~(\ref{EQ}) considering small masses $ \epsilon\, m^2$ instead of $m^2$.
Eq.~(\ref{EQBFKL}) can then , be re-written with a different way of regularization, for example
\begin{equation} \label{EQBFKL0}
E \,\phi^{\mbox{\tiny BFKL}}_E\left( \kappa\right)\,\,\underset{\epsilon\,\to\,0}{=}\,\,- \ln \left( \frac{\epsilon^2}{2}\right)\, \phi^{\mbox{\tiny BFKL}}_E\,\,-\,\,\int^{\infty}_{0}\,d \kappa' \,\phi^{\mbox{\tiny BFKL}}_E\left( \kappa'\right)\,\frac{\Theta\left( |\kappa - \kappa'| - \epsilon\right)}{|\kappa\,-\,\kappa'|\,+\,\epsilon},
\end{equation}
where $\Theta\left( z \right)$ is the unit step function which is equal 1 for $z > 0$ and 0 for $z< 0$. At large $\kappa$ solutions of Eq.~(\ref{EQF}),  should coincide with the eigenfunctions of Eq.~(\ref{EQBFKL}) which 
are well known~~\cite{BFKL,LIREV} and have the form
\begin{equation} \label{LKAP}
\phi_E\left( \kappa\right)\,\,\xrightarrow{\kappa \to \infty}\,\,\phi^{\mbox{\tiny BFKL}}_E\left( \kappa\right)\,\sim\,\kappa^{\gamma -1}
\,\,\,\,\,\,\mbox{with}\,\,\,\,\,\,E\left( \gamma \right)\,\,=\,\,-\chi\left( \gamma\right)\,\,=\,\,\psi\left( \gamma\right)\,+\,\psi\left( 1 - \gamma\right) \,-\,2 \psi\left( 1 \right),
\end{equation}
where $\psi(x)=\Gamma'(x)/\Gamma(x)$ is a digamma function. Two eigenfunctions, $\phi_E\left( \kappa\right)\,\propto\,\kappa^{\gamma -1}$ and $\phi_E\left( \kappa\right)\,\propto\,\kappa^{- \gamma }$, describe the states with the same energy. For $\gamma = \frac{1}{2} + i \nu$ these eigenfunctions are  normalized and form a complete set of functions.

 \section{Semi-classical approach: generalities and equations}
 
 \subsection{The main qualitative features of the solution}
  For  completeness of presentation, we start discussing the solutions to Eq.~(\ref{EQF}), repeating the key qualitative and general features of  solutions that have been discussed in Ref.~\cite{LLS}. The first one has been mentioned in the previous section: at large values of $\kappa$, 
  the eigenfunctions $\phi_E\left( \kappa\right)$ should approach the eigenfunctions of the massless BFKL equations~(\ref{LKAP}).
  
  The behavior of the solutions at small values of $\kappa$,   is easier to understand by rewriting Eq.~(\ref{EQF}) in the coordinate representation. Using 
\begin{equation} \label{CORE1}
\int \frac{d^2 p'}{2 \pi} \frac{e^{i \vec{r} \cdot \vec{p}^{\,\,'}}}{ p'^2 + m^2}\,=\,\int^{+ \infty}_{-\infty}\frac{p' d p'\,J_0\left( r p'\right)}{p'^2 + m^2 }\,\,=\,\,K_0\left( r m\right)
\end{equation}
where $J_0\left( z\right)$ and $K_0\left( z\right)$ are the Bessel and MacDonald functions~~\cite{RY}, we can re-write Eq.~(\ref{EQF}) in the form
\begin{equation} \label{H}
E\,f\left( r \right)\,\,=\,\,{\cal H}\,f_E\left( r \right),~~~\mbox{where}~~~ f_E\left( r \right)\,\,=\,\,\int \frac{d^2 p}{(2 \pi)^2}\,e^{ i \vec{p} \cdot \vec{r}}\,\phi_E\left( \frac{p^2}{m^2} \right),
\end{equation}
and
\begin{eqnarray}\label{H1}
{\cal H}\,&=&\,\frac{p^2 +m^2}{|p|\sqrt{p^2 + 4  m^2}}\ln \frac{\sqrt{p^2 + 4 m^2} + |p|}{\sqrt{p^2 + 4 m^2} - |p|}\,\,-\,\,2 K_0\left( |r| m \right)\,+\,\frac{N^2_c + 1}{2\,N^2_c} \hat{P}\,\,\nn\\
&=&\,\,T\left( p^2 \right) \,\,+\,\,V\left( r \right) \,\,+\,\,\frac{N^2_c + 1}{2\,N^2_c} \hat{P}\,\,=\,\,{\cal H}_0\,\,+\,
\,\,\frac{N^2_c + 1}{2\,N^2_c} \hat{P}.\end{eqnarray}
In~(\ref{H1}) we introduced a shorthand notation $\hat{P}$ for the projector onto the state $\sim m^2/(p^2+m^2)$
\begin{equation} \label{P}
\hat{P} \,\phi\left( p \right)\,=\,\frac{m^2}{p^2 + m^2}\int \frac{d^2 p'}{\pi} \frac{\phi\left( p'\right)}{p'^2\,+\,m^2}\,\,\,\xrightarrow{\mbox{in coordinate representation}}\,\,\,\, K_0\left( |r| m\right) \int \frac{d^2 p'}{\pi} \frac{\phi\left( p'\right)}{p'^2\,+\,m^2}
\end{equation}
 
 The behavior of Eq.~(\ref{LKAP}) at large $p$,  translates into the following behavior at short distances
 \begin{equation} \label{CORE2}
  f_{E\,=\,E\left( \ga\right)}\left( r \right) \,\,\,\,\xrightarrow{r \ll 1/m}\,\,\,f^{\mbox{\tiny BFKL}}_{E\,\,=\,\,E\left( \ga\right)}\left( r\right)\,\,\sim\,\,\left( r^2\right)^{\gamma -1  } \,\cup\,\left( r^2\right)^{- \gamma   }   \end{equation}
  To understand the behavior of the solutions at large distances, we should distinguish the two cases, when the wave function's decrease is   slower than $e^{-mr}$ , and when  the decreases are faster $e^{-mr}$. In the former case, as we may see from~Eq.~(\ref{H1},\ref{P}), we may neglect the contact term and term $\sim V(r)$, and Eq.~(\ref{H}) degenerates into
  \begin{equation} \label{CORE3}
  T  \, f^0_{E\,=\,E\left( \ga\right)}\left( r \right)  \,\,=\,\,\frac{\kappa + 1}{\sqrt{\kappa}\sqrt{\kappa + 4  }}\ln \frac{\sqrt{\kappa + 4 } + \sqrt{\kappa}}{\sqrt{\kappa + 4 } - \sqrt{\kappa}}    \, f^0_{E\,=\,E\left( \ga\right)}\left( r \right)\,\,=  
  \,\,E\left( \nu \right)\, f^0_{E\,=\,E\left( \ga\right)}\left( r \right)  
  \end{equation}
  
  where $\kappa = - \nabla^2_r$.
  
  The eigenfunctions of Eq.~(\ref{CORE3}) have a form
\begin{equation} \label{CORE4}
 f^0_{E\,=\,E\left( \nu\right)}\left( \vec{r}\right)\,\,\sim\,\, e^{ i \sqrt{- a} r}~~\mbox{for}~~a \,<\,0;~~~\mbox{and}~~~ f^0_{E\,=\,E\left( \nu\right)}\left( \vec{r}\right)\,\,\sim\,\, e^{ - \sqrt{ a} r}~~\mbox{for}~~a \,>\,0; \end{equation}
 
\begin{figure}[ht]
     \begin{tabular}{c c}
     \includegraphics[width=8cm]{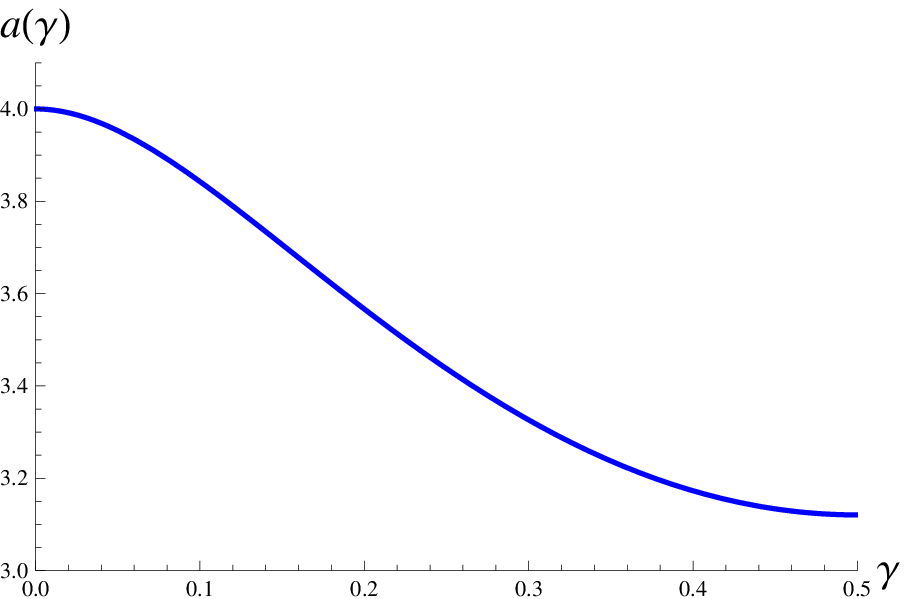} &   \includegraphics[width=8cm]{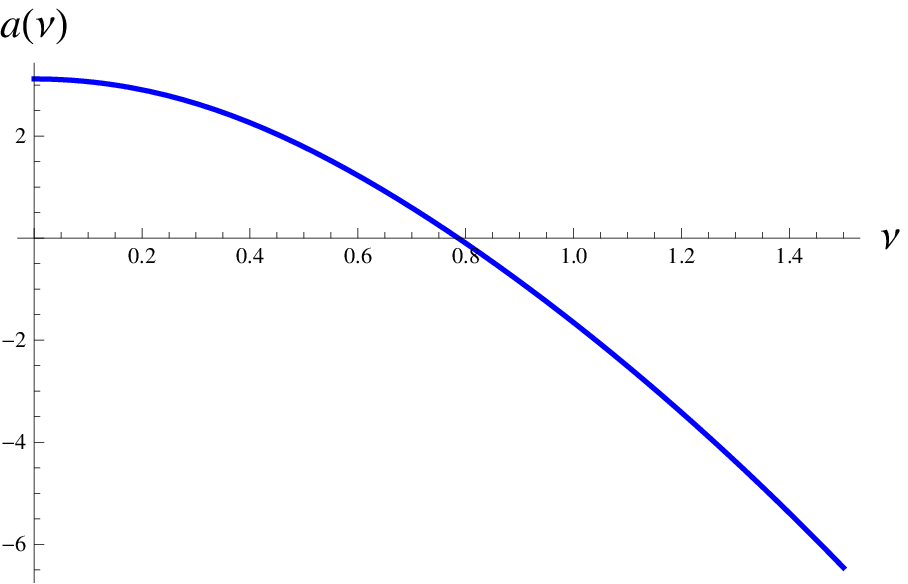}  \\
      Fig.~\protect\ref{a}-a &  Fig.~\protect\ref{a}-b
        \end{tabular}    
      \caption{ Function $a\left( \ga \right)$ (solution to Eq.~(\protect\ref{CORE5})) versus $\ga$ (\protect Fig.~\protect\ref{a}-a) and versus $\nu$ where $\gamma = \frac{1}{2}\, + \,i \,\nu$ (\protect Fig.~\protect\ref{a}-b).  }
\label{a}
   \end{figure}

 From~Eq.~(\ref{CORE3}) we can see that parameters $a$ and $\ga$ are correlated, viz.
 \begin{equation} \label{CORE5} 
E\,\,=\,\,T\left( - a\right)\,\,=\,\,- \chi\left( \ga\right)
 \end{equation}
 The solution to this equation is shown in  Fig.~\protect\ref{a}.
 The Fourier image of $1/(p^2 + m^2)^{1 - \ga}$  in the coordinate representation is
 \begin{equation} \label{FRRIM}
 \frac{1}{(\kappa + a)^{1 - \ga}}\xrightarrow{\mbox{\tiny Fourier image}}\frac{1}{\Ga\left( 1 - \ga\right)}\left( \frac{2 \sqrt{a}}{r}\right)^\ga \,K_\gamma\left( \sqrt{a} r\right) \,\xrightarrow{\sqrt{a} r \gg 1}\,
 \frac{1}{\Ga\left( 1 - \ga\right)}\left( \frac{2 \sqrt{a}}{r}\right)^\ga\sqrt{\frac{\pi}{2 \sqrt{a} r}}\,e^{- \sqrt{a} r}
 \end{equation}
The   function $1/(\kappa + a)^{1 - \ga}$ describes both the short distance~(\ref{CORE2}) and long distance~(\ref{CORE4}) behaviors. In~\cite{LLS} we demonstrated that the ground state with the minimal energy is  reached with this class of functions. 
 \subsection{ Semi-classical approach: generalities}
  \subsubsection{ The method of steepest descent} 
  \label{SDM}
For massless BFKL, a general solution has a form
\begin{equation} \label{MT:massless}
\Psi\left( Y,  \kappa \right)\,\,=\,\,\int^{\epsilon + i \infty}_{\epsilon - i \infty}\frac{d \ga}{2 \pi i} \,\phi_{in}\left( \ga\right)\,e^{ - E\left( \ga\right)\,Y + \left( \ga - 1\right) \ln \kappa},
\end{equation}
 where $\phi_{in}\left( \ga\right)$ should be found from the initial conditions at $Y = 0$. For the  massive case, we will look for solutions of~Eq.~(\ref{EQY}) of the  analogous form
\begin{equation} \label{MT}
\Psi\left( Y, l \right)\,\,=\,\,\int^{\epsilon + i \infty}_{\epsilon - i \infty}\frac{d \ga}{2 \pi i} \,\phi_{in}\left( \ga\right)\,e^{ - E\left( \ga, l\right)\,Y + \left( \ga - 1\right) l},
\end{equation}
 where we introduced a new variable $l\,\,=\,\,\ln\left( \kappa + a\right)$, which is stable in the small-$\kappa$ limit, and $\phi_{in}\left( \ga\right)$ is fixed by the initial conditions at $Y = 0$. In Eq.~(\ref{MT}) we can take the integral over $\ga$ using the method of steepest descent, which is equivalent to a search of semiclassical solution to Eq.~(\ref{EQY}). The equation for the saddle point takes the general form
 \begin{equation} \label{SP}
 - \frac{ \partial E\left( \ga_{SP}\left( Y, l \right) , l\right) }{\partial \ga}\,Y\,+\, l \,=\,0.
 \end{equation}
 If $\phi_{in}\left( \ga \right)$ in the integrand of~(\ref{MT}) is a smooth function, we can replace it with its value at $\ga = \ga_{SP}$.
 The integral over $\gamma$ in the vicinity of a saddle-point~(\ref{SP}) yields
 \begin{eqnarray} \label{SP1}
&& \Psi\left( Y, l \right)\,\,=\\
 && \,\,\phi_{in}\left( \ga_{SP}\left( Y, l\right)\right)\,e^{ - E\left( \ga_{SP}\left( Y, l\right), l\right)\,Y + \left( \ga _{SP}\left( Y, l\right)- 1\right) l}\int^{\epsilon + i \infty}_{\epsilon - i \infty}\frac{d \ga}{2 \pi i} \, \exp\Big( - \frac{1}{2}  \frac{\partial^2 E\left( \ga_{SP}\left( Y, l \right) , l\right) }{\partial \ga_{SP}^2}\,Y \left( \ga - \ga_{SP}\left( Y, l\right) \right)^2\Big) \nn\\
 && =   \phi_{in}\left( \ga_{SP}\left( Y, l\right)\right)\,\sqrt{\frac{1}{ 2 \pi \, \Big{|}\frac{\partial^2 E\left( \ga_{SP}\left( Y, l \right) , l\right)}{\partial \ga^2_{SP}}\Big{|}\, Y}}\,\, e^{S\left( Y, l \right)}\,\,=\,\, \phi_{in}\left( \ga_{SP}\left( Y, l\right)\right)\,\sqrt{\frac{1}{ 2 \pi \, \Big{|}\frac{\partial^2 E\left( \ga_{SP}\left( Y, l \right) , l\right)}{\partial \ga^2_{SP}}\Big{|}\, Y}}\,\, e^{-\frac{1}{2} l}\,e^{\omega_{eff}\left( l, Y \right) \,Y}.\nn
   \end{eqnarray}
 The omission of higher-doer corrections in Eq.~(\ref{SP1}) is justified  due to  smallness of the parameter
 \begin{equation} \label{SCCON}
R\,\,=\,\, \frac{1}{6}\Big{|}\frac{\partial^3 E\left( \ga_{SP}\left( Y, l \right) , l\right)}{\partial \ga_{SP}^3}\Big{|}\,Y\Bigg{/}\Big( \frac{1}{2} \Big{|}\frac{\partial^2 E\left( \ga_{SP}\left( Y, l \right) , l\right)}{\partial \ga^2_{SP}}\Big{|}\, Y\Big)^{3/2}\ll 1.
 \end{equation}
   The  smoothness of the initial function $\phi_{in}\left( \ga\right)$ implies a condition
   \begin{equation} \label{SCCONPHI}
\frac{1}{\sqrt{\Big{|} \frac{1}{2}\frac{ \partial^2 E\left( \ga_{SP}\left( Y, l \right) , l\right)}{\partial \ga^2_{SP}}\Big{|}\, Y}}\,\ll\,\frac{\ln \phi_{in}\left( \ga\right)}{d \ga}|_{\ga = \ga_{SP}}
  \end{equation} 
    and determines the kinematic region of applicability of the semiclassical approximation. As we will demonstrate below,  both conditions are satisfied for sufficiently large $Y \,\gg\,1$.

  \subsubsection{Solution with method of characteristics}
\label{subsec:MC}
 The method of characteristics for a partial differential equation (PDE) corresponds to a reduction of the PDEs 
 \begin{equation}
  F(x_1,\dots,x_n,u,p_1,\dots,p_n)=0,\qquad p_i=\frac{\partial u}{\partial x_i}
 \end{equation}
 to a system of ordinary differential equations (ODE) for characteristic lines along which the PDE converts into an ordinary differential equation. These characteristics satisfy the Lagrange-Charpit equations
 \begin{equation}
    \frac{\dot{x}_i}{F_{p_i}}=-\frac{\dot{p}_i}{F_{x_i}+F_up_i}=\frac{\dot{u}}{\sum p_iF_{p_i}}. 
 \end{equation}
 
 A n instructive  example familiar from a classical mechanics, is a Hamilton-Jacobi equation, for which the characteristic lines correspond to trajectories of particles which are solutions of the Newtonian  equations of motion. A detailed discussion of the method is beyond the scope of the present paper and can  be found in the literature (see e.g. textbooks~~\cite{Fritz,Kamke}).
 
 A direct application of the method of characteristics to the evolution equation Eq.~(\ref{EQY}) is not straightforward , since it is \textit{integro}differential equation . However, as we will show below, for  a special case which corresponds to a semiclassical approximation, this method is applicable. It is convenient to rewrite the wave function $\Psi$ in terms of the ``action'' $S$~~\cite{FeynmanHibbs},
 
 \begin{equation}
  \Psi(Y,\kappa)=e^{S(Y,\kappa)}.
 \end{equation}
 
 Then the evolution equation~Eq.~(\ref{EQY}) takes the form of a nonlinear PDE
\begin{equation} \label{EQMC}
 F\left( Y,l, S,\,\gamma,\omega\right)\,\,=\,\,\om\left( Y, l\right)  \,-\, \chi\left(\ga\left( Y,l\right) \right)\,\,-\,\,\widetilde{P}\left( l,\ga\left( Y, l,a\right)\right)  \,\,=\,\,0
 \end{equation}
 
 where we introduced shorthand notations

\begin{equation} \label{MC2} 
 \om\left( Y,l\right)\,=\,\frac{\D S\left( Y; l\right)}{\D Y};~~~~\ga\left( Y,l\right)\,-\,1\,=\,\frac{\D S\left( Y; l\right)}{\D l}~~~
  \end{equation}
and introduced a new variable $l\,\,=\,\,\ln\left( \kappa\,+\,a\right)$ which remains finite in the small-$\kappa$ limit. The explicit form of the function $\widetilde{P}\left( l,\ga\left( Y, l,a\right)\right)$ in~Eq.~(\ref{MC2}) will be specified below in section~\ref{SCAKBEWMG}, here we will only mention that $\widetilde{P}\left( l,\ga\left( Y, l,a\right)\right)\,\,\xrightarrow{ l \,\gg\,1} \,\,0$. From the exact solutions of the massless BFKL~\ref{LKAP}, we expect that at large $\kappa$, the effective action should  depend linearly on rapidity $Y$ and a new variable $l$, 
\begin{equation}
 S_{SC}\approx \omega_\infty + (\gamma_\infty-1)l,
\end{equation}

where $\gamma_\infty$ and $\omega_\infty=\omega_{\rm BFKL}(\gamma_\infty)$ are constants. In a semiclassical approximation which is valid for the moderate values of $Y$, we assume that\footnote{We introduce a subscript index $SC$ for semiclassical approximation} $\omega_{SC}$ and $\gamma_{SC}$ are slowly varying functions of variables $l$, $Y$: viz. $\omega'_Y\Lb  Y, l \Rb \,\ll\,\omega^2\Lb  Y, l \Rb$, $\omega'_l\Lb  Y, l \Rb \,\ll\,\omega^2\Lb  Y, l \Rb$ and $ \gamma'_Y\Lb  Y, l \Rb \,\ll\,\Lb 1 - \gamma\Lb  Y, l \Rb\Rb^2$, 
$ \gamma'_l\Lb  Y, l \Rb \,\ll\,\Lb 1 - \gamma\Lb  Y, l \Rb\Rb^2$.  Making  this assumption, the equation~Eq.~(\ref{EQMC}) has a form of the PDE which can be solved using the method of characteristics~~\cite{Fritz,Kamke}. The characteristic lines  $l(t), Y(t), S(t)$, $ \omega(t)$ and $\gamma_{SC}(t)$, where $t$ is some parameter (analog of time in case of classical mechanics), satisfy a system of ODEs:
\begin{eqnarray}
&&\hspace{-0.3cm}(1.)\,\,\,\,\frac{d l}{d\,t}\,\,=\,\,\frac{\partial F}{\partial \gamma}\,\,= \,\,- \frac{d \chi\left( \ga_{SC} \right)}{d \ga_{SC}} \,\,-\,\,\frac{\D \widetilde{P}\left( l(t), \ga_{SC}(t), a\right)}{\D \ga_{SC}}\label{MC4}\\
&&\hspace{-0.3cm}(2.)\,\,\,\,\,\,\,
\frac{d\,Y\left( t\right)}{d\,t}\,\,=\,\,\frac{\partial F}{\partial \omega}\,\,=\,\,1\,\,\,\,\,\,\,\nn\\
 &&\hspace{-0.3cm}(3.)\,\,\,\,
\frac{d S}{d\,t}\,\,=\,\,\left(\gamma - 1\right)\,\frac{\partial F}{\partial \gamma}\,+\,\omega\,\frac{\partial F}{\partial \omega}\,\,=\,\,  \,\,\left(\ga_{SC} - 1\right)\Big\{- \frac{d \chi\left( \ga_{SC} \right)}{d \ga} \,\,-\,\,\frac{\D \widetilde{P}\left( l(t), \ga_{SC}(t), a \right)}{\D \ga}\Big\}\,\,+\,\,\om_{SC}
\nn\\
&&\hspace{-0.3cm}(4.)\,\,\,\,\frac{d\,\gamma_{SC}}{d\,t}\,\,=\,\,-
\left(\,\frac{\partial F}{\partial l}\,+\,\left( \gamma_{SC}\,-\,1\right)\,\frac{\partial F}{\partial S}\,\right)\,\,=\,\,  \frac{\D \widetilde{P}\left( l(t),\ga_{SC}(t),a\right)}{\D\, l}\nn\\
&&\hspace{-0.3cm}(5.)\,\,\,\,\frac{d\,\omega_{SC}}{d\,t}\,\,=\,\,-
\left(\,\frac{\partial F}{\partial Y}\,+\,\omega \,\frac{\partial F}{\partial S}\,\right)\,\,=\,\,  0\nn
\end{eqnarray}
The second line of equation~Eq.~(\ref{MC4}) implies that the parameter $t$ corresponds to a rapidity $Y$. From the fifth line of~Eq.~(\ref{MC4}) we can see that $\omega_{SC}$ is conserved on characteristic lines, and can be fixed from the asymptotic conditions~Eq.~(\ref{LKAP}) as
  \begin{equation} \label{MC5}
  \omega_{SC}\,\,=\,\,\chi\left( \gamma_{SC}\right)\,\,+\,\,\widetilde{P}\left( l(t),\ga_{SC}(t),a\left( \gamma_\infty\right)\right)\,\,=\,\,\chi\left( \gamma_{\infty}\right).   \end{equation} 
    
A combination of Eq.~(\ref{MC4})-4 and Eq.~(\ref{MC4})-1 allows us to eliminate  the $Y$-dependence,  and find $\gamma_{SC}$ as a function of $l$ on a trajectory,
\begin{equation} \label{MC6}
\frac{d\,\gamma_{SC}}{d\,l}\,\,=\,\,-  \frac{\D \widetilde{P}\left( l(t),\ga_{SC}(t), a\right)}{\D\, l}\Bigg{/}\left( \frac{d \chi\left( \ga_{SC} \right)}{d \ga_{SC}} \,\,+\,\,\frac{\D \widetilde{P}\left( l(t), \ga_{SC}(t), a\right)}{\D \ga_{SC}}\right)
\end{equation}

Using  Eq.~(\ref{MC4})-1 we can write the equation for  the trajectory  $l = l\left( Y\right)$, which has the form
\begin{equation} \label{MC7}
\frac{d l\left( Y\right)}{d Y}\,\,=\,\,\,\,- \frac{d \chi\left( \ga_{SC}\left( l\left( Y\right)\right) \right)}{d \ga} \,\,-\,\,\frac{\D \widetilde{P}\left( l(Y), \ga_{SC}(l\left( Y\right)), a\right)}{\D \ga_{SC}}
\end{equation}

The lines  $ l \left( Y \right) \,\equiv\, l_{SC}\left( Y\right)$ give the set of trajectories.  The trajectory that leads to the dominant contribution to $\Psi\left( Y, l\right)$ can be found from the equation
\begin{equation} \label{MC71}
l\,=\,l_{SC}\left( Y; \gamma_{\infty}\right)\,\,=\,\,l_{SP}\left( Y; \gamma_{\infty}\right)\end{equation}
Corresponding $\gamma_{SC}\left( Y, l_{SC}\left( Y\right); \gamma_{\infty}\right)\,=\,\gamma_{SP}\left( Y, l_{SP}\left( Y\right); \gamma_{\infty}\right)$.
Finally, Eq.~(\ref{MC4})-3 can be re-written in the form
\begin{equation} \label{MC40}
  \frac{d S}{d\,Y}\,\,=\,\,\left(\gamma_{SC}\left( Y, l \right) - 1\right)\,\frac{d l\left( Y \right)}{d Y}\,\,+\,\, \chi\left(\ga_{SC}\left( Y,l\right) \right)\,\,+\,\,\widetilde{P}\left( l,\ga_{SC}\left( Y, l,a\right)\right)  
  \end{equation} 
  
 Comparing Eq.~(\ref{MC4}) with Eq.~(\ref{SP1}) one can see that the effective intercept is equal to
 $\omega_{eff}\left( Y, l\right)\,\,=\,\,\Big(S\left( Y, l_{SP}\left( Y\right)\right)\,+ \,\frac{1}{2} \,l_{SP}\left( Y\right)\Big)/ Y$.

Before applying the semiclassical approach to the BFKL equation for massive gluons, in the following subsection~\ref{BFKL:massless}, we would like to test it on  the massless BFKL equation~(\ref{EQBFKL}), for which analytical solutions are known~(see for example Ref.~\cite{GLR,REV}), and only after that in subsection~\ref{SCAKBEWMG} we apply it to the massive case.
  
 \subsection{ Semi-classical approach:  massless BFKL equation}
\label{BFKL:massless}
\begin{figure}[ht]
     \begin{tabular}{c c c}
     \includegraphics[width=5.3cm]{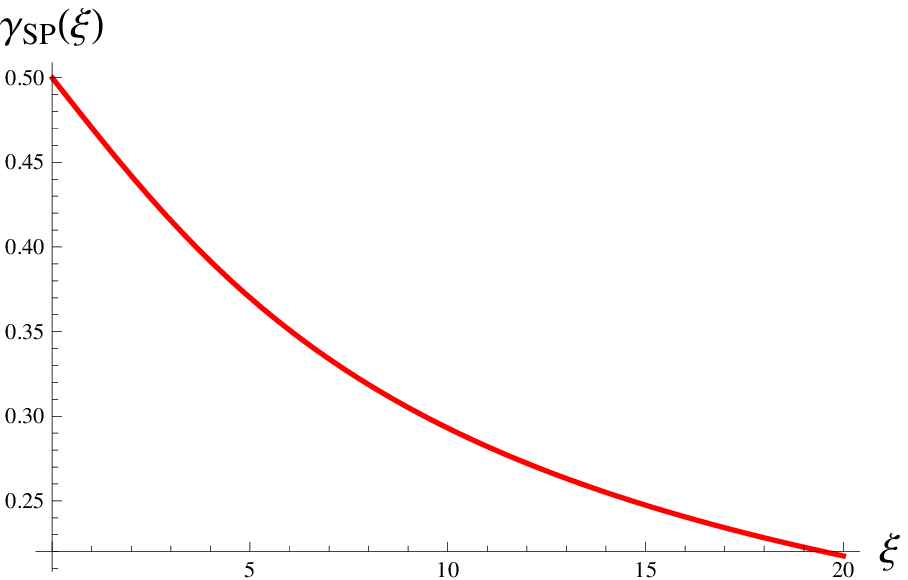} & \includegraphics[width=5.3cm]{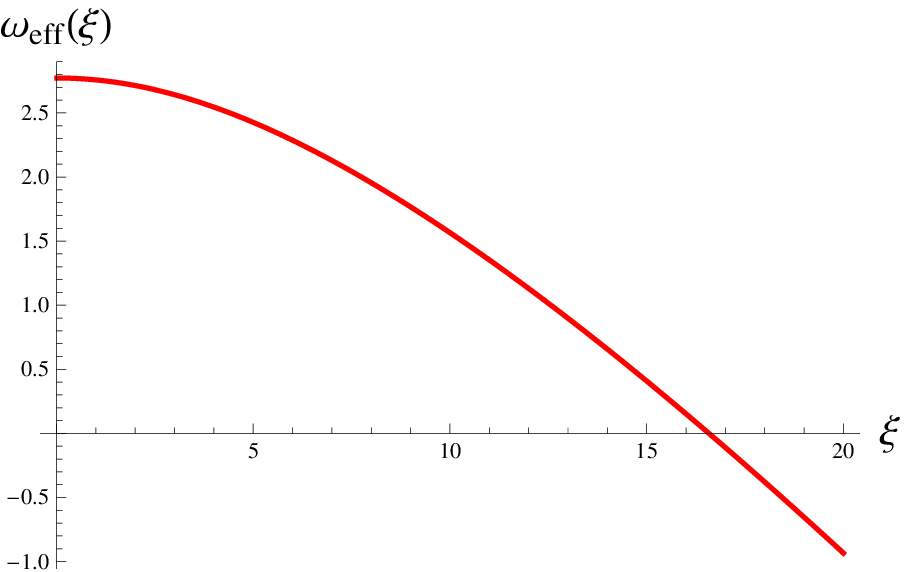}
     & \includegraphics[width=5.3cm]{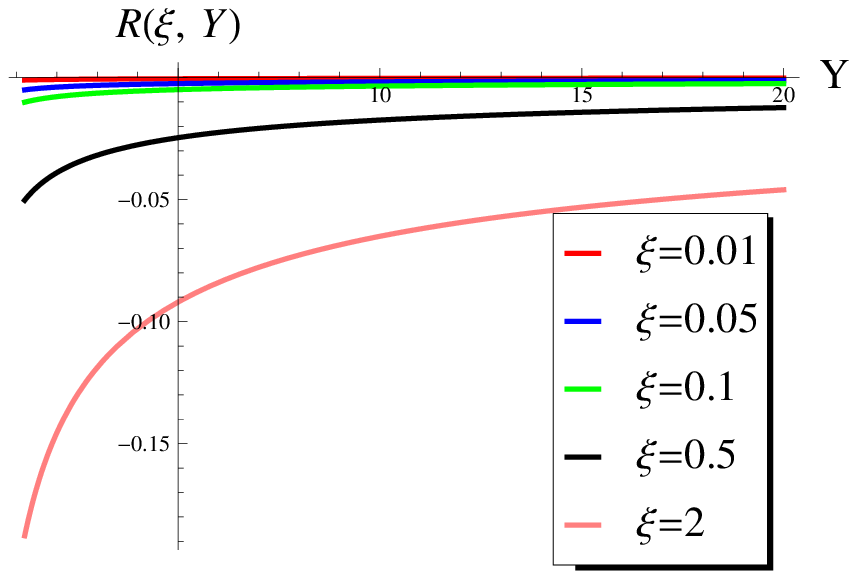}\\
      Fig.~\protect\ref{m0}-a & Fig.~\protect\ref{m0}-b & Fig.~\protect\ref{m0}-c\\
      \end{tabular}    
      \caption{ Semi-classical solution for massless BFKL equation( see  Eq.~(\protect\ref{EQBFKL})):
      trajectories as function of $\xi = l/Y$ (\protect Fig.~\protect\ref{m0}-a); effective Pomeron intercept ($\omega_{eff}$  versus $\xi$ (\protect Fig.~\protect\ref{m0}-b); and the ratio $R$ of  Eq.~(\protect\ref{SCCON}) at fixed $\xi$ as a function of $Y$(\protect Fig.~\protect\ref{m0}-c).}
\label{m0}
   \end{figure}

For Eq.~(\ref{EQBFKL}) $ E\left( \ga_{SP}\left( Y, l \right) , l\right)\,=\,- \chi\left( \ga \right)$ (see Eq.~(\ref{LKAP})).  Combining this equation with Eq.~(\ref{SP}), we obtain for  the saddle point $\ga_{SP}\left( Y,l = \ln \kappa\right)$
\begin{equation} \label{SCM0}
\chi'_\ga\left( \ga_{SP}\left( \xi\right)\right)\,+\,\xi =0~~~~~~~\mbox{where}~~~~~~~\xi \,=\,l/Y.
\end{equation}
The solution to Eq.~(\ref{SCM0}) is shown in  Fig.~\protect\ref{m0}-a. From Eq.~(\ref{SP1}) one has that 
\begin{equation} \label{SCM01}
 \Psi_{\mbox{\tiny BFKL}}\left( Y, l  = \ln \kappa \right)\,\,\propto \,\,e^{\omega_{eff}\left( \xi\right) Y},
 ~~~~\mbox{where}~~~~~\omega_{eff}\left( \xi\right)\,\,=\,\,\chi\left( \ga_{SP}\left( \xi\right)\right) \,+\,\left( \ga_{SP}\left( \xi\right) - \frac{1}{2} \right) \xi.
 \end{equation}
 The dependence $\omega_{eff}\left( \xi\right)$ is plotted in  Fig.~\protect\ref{m0}-b. The Fig.~\protect\ref{m0}-c shows the ratio $R\left( \xi,Y\right)$ defined in Eq.~(\ref{SCCON}). This ratio is small at large values of $Y$ and small $\xi$, thus justifying the applicability of semiclassical approximation in this region.
 \subsubsection{ Diffusion approximation}
\label{sub:DiffApprox}
As one can see from  Fig.~\protect\ref{m0}-a at $\xi \to 0$ or in other words, at large $Y \gg l$, the trajectory $\ga_{SP}$ approaches a limiting value $\ga_{SP}(\infty) = \frac{1}{2}$. 
Since $\chi(\gamma)$ is an analytic function near this point, it can be approximated by
\begin{equation} \label{DIFA}
 \chi\left(\ga\right)  \,\,\,=\,\,\,\omega_0\,\,+\,\,D\,\left( \ga - \frac{1}{2}\right)^2\,\,+\,\,{\cal O}\left( \left( \ga - \frac{1}{2}\right)^3\right),
\end{equation}
   with $\omega_0 = 4 \ln 2 = 2.772$\,\,\,\mbox{and}\,\,\,$D = 14 \zeta(3) = 16.822$. In the approximation~(\ref{DIFA}) the Eq.~(\ref{SCM0}) can be solved  easily and yields
   \begin{equation} \label{DIFA1}
   \ga_{SP} = \frac{1}{2} - \frac{\xi}{2 D}.
   \end{equation}
   Substituting (\ref{DIFA1}) into Eq.~(\ref{SP1}), we obtain
   \begin{equation} \label{DIFA2}
    \Psi{\mbox{\tiny BFKL}}\left( \xi , Y\right)\,=\, \phi_{in}\left( \frac{1}{2} - \frac{\xi}{2 D}\right)\,\sqrt{\frac{1}{ 4 \pi \,D \, Y}}\,\,e^{- \frac{1}{2} l} e^{\omega_0 Y  - \frac{\xi^2}{4 D Y} },
    \end{equation}
    \textit{i.e.}, the semiclassical approach reproduces the diffusion approximation~\cite{REV} for the BFKL equation.
 \subsubsection{ Double log  approximation}
For small values of $\ga \approx 0$, the BFKL kernel can be approximated by
\begin{equation} \label{DLA}
 \chi\left(\ga\right)  \,\,\,=\,\,\,\frac{1}{\ga}.
 \end{equation}
 In this limit, Eq.~(\ref{SCM0}) gives the trajectory  $\ga_{SP} = 1/\sqrt{\xi}$,  and the wave function
 \begin{equation} \label{DLA1}
    \Psi_{\mbox{\tiny BFKL}}\left( \xi , Y\right)\,=\, \phi_{in}\left( 1/\xi\right)\,\sqrt{\frac{1}{ 2 \pi \,\xi^{3/2} \, Y}}\,\,e^{2 \sqrt{\xi} Y}\,\,=\,\,\phi_{in}\left( Y/l\right)\,\sqrt{\frac{1}{ 2 \pi \,l^{3/2} \, Y^{-1/2}}}\,\,e^{2 \sqrt{Y\,l} }.
    \end{equation}
     Eq.~(\ref{DLA1}) is the solution in the double log approximation~\cite{REV}.

 \subsection{ Semi-classical approach: kernels of the BFKL equation with massive gluon}
  \label{SCAKBEWMG}
\begin{figure}[ht]
     \begin{tabular}{c c c}
     \includegraphics[width=5.5cm]{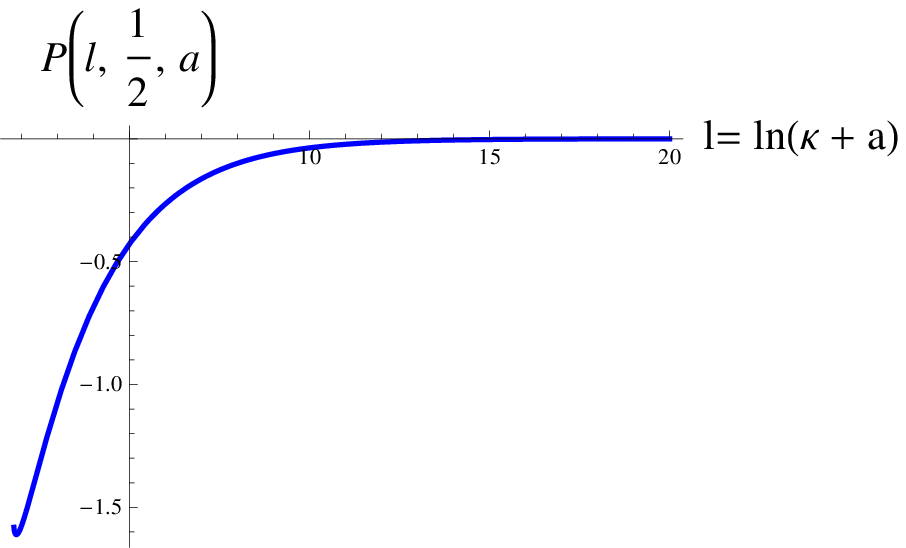} &     \includegraphics[width=5.5cm]{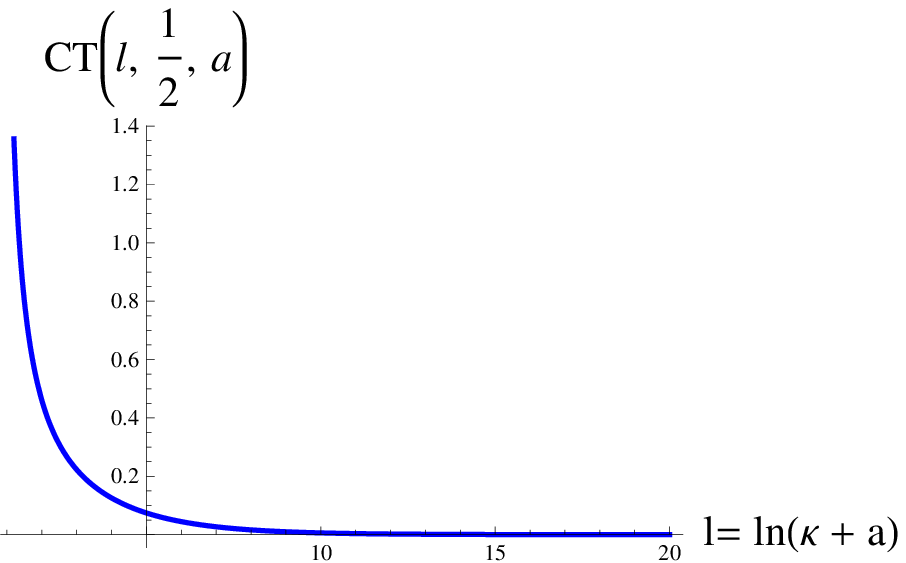}& \includegraphics[width=5.5cm]{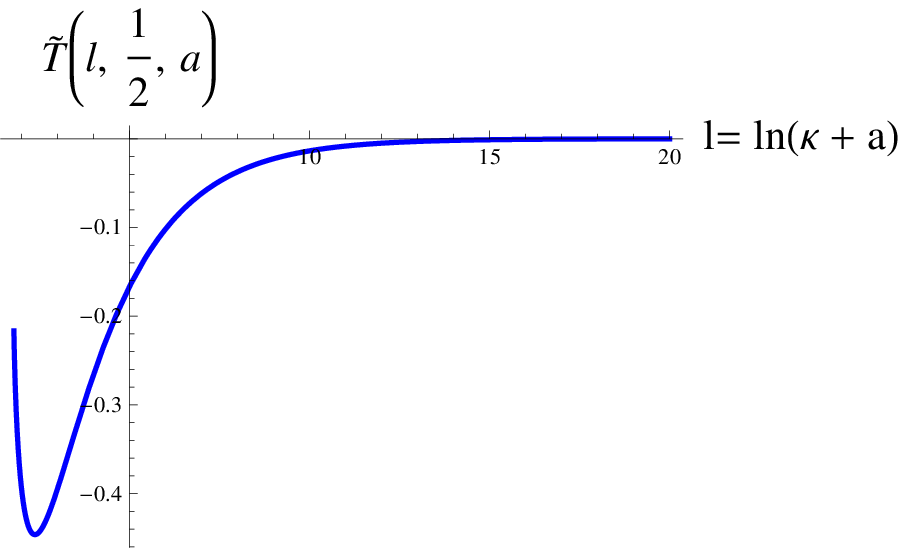}  \\
      Fig.~\protect\ref{ptct}-a & Fig.~\protect\ref{ptct}-b & Fig.~\protect\ref{ptct}-c \\
     \end{tabular}     
      \caption{All ingredients of  Eq.~(\protect\ref{ptct})  versus $l$ at $\gamma = \frac{1}{2}$ and  $a = a\left( \frac{1}{2}\right)$ ( see 
         Eq.~(\protect\ref{CORE5}) and \protect Fig.~\protect\ref{a}). }
\label{ptct}
   \end{figure}

  Plugging the solution Eq.~(\ref{MT}) into Eq.~(\ref{EQY}), we obtain the equation for $E\left( \ga, l \right)$ in the form 
  \begin{eqnarray} \label{SC2}
   E\left( \ga, l \right)\,\,&=&\,\,T\left( e^l - a \right)\,\,+\,\,CT\left( l,\gamma\left( l \right),a;\right)\,\,-\,\,{\cal K}\left( l, \gamma\left( l \right), a\right),
  \end{eqnarray}
  where the emission kernel ${\cal K}\left( l, \gamma\left( l \right), a\right)$ is given by
  \begin{equation} \label{SC3}
  {\cal K}\left( l, \gamma\left( l \right), a\right) \,\,=\,\,\int^1_{\frac{a}{\kappa}} \frac{t^{\gamma\left( l \right) - 1} \,d t}{
  \sqrt{ (1 - t)^2 \,+\,\frac{2}{ \kappa}( 1 + t) \,\,+\,\,\frac{1 - 4 a}{\kappa^2}}}
  \,\,+\,\, \int^1_{0} \frac{t^{- \gamma\left( l \right)} \,d t}{
  \sqrt{ (1 - t)^2 \,+\,\frac{2}{ \kappa} t ( 1 + t) \,\,+\,\,( 1 - 4 a)\frac{t^2}{ \kappa^2}}}.
  \end{equation}
 The integral over $t$ in Eq.~(\ref{SC3})  can be evaluated analytically and expressed in terms of the Appel function $F_1$\footnote{See Eqns. {\bf 9.180 - 9.184} in Ref.~\cite{RY}}:   
\begin{eqnarray}\label{SC5}
{\cal K}\left( l, \gamma\left( l \right), a\right)&=&2 \sqrt{\frac{t^0 - t^+}{t^+ - t^-}} \left( t^+\right)^{1 - \gamma\left( l \right)} F_1\left( \frac{1}{2}, 1 - \gamma\left( l \right), \frac{1}{2}, \frac{1}{2} ,1 - \frac{t^0}{t^+}, \frac{t^+ - t^0}{t^+ - t^-}\right)\\
 &+& \sqrt{\pi}\,\,\frac{\Gamma\left( 
 \gamma\left( l \right)\right)}{\Gamma\left( \frac{1}{2}
  + \gamma\left( l \right)\right)}{}_2F_1\left( \frac{1}{2}, \frac{1}{2},  \gamma\left( l \right)  - \frac{1}{2}, \frac{t^+}{t^+ - t^-}\right)\,\nn\\
  &+&\,\,B\left( 1, \gamma\left( l \right)\right)\,F_1\left( \gamma\left( l \right), \frac{1}{2}, \frac{1}{2}, 1 +  \gamma\left( l \right), t^-, t^+\right),\nn
 \end{eqnarray}
 where 
 \begin{equation} \label{SC6}
 t^0\,\,=\,\,\frac{a}{ \kappa};~~~~~ t^{\pm}\,\,=\,\, 1 \,+\,\frac{1}{ \kappa}\,\,\,\pm\,\,\,2\, \sqrt{\frac{t^0\,-\,1}{ \kappa}}.
 \end{equation}
 For practical reasons it is convenient to introduce a function $P\left( l, \ga, a \right)$ defined as 
\begin{eqnarray} \label{SC7}
P\left(  l, \ga, a  \right) &=& \int^1_{t^0} d t  \left( t^{\ga - 1} - 1\right) \Big[\frac{1}{\sqrt{( 1 - t)^2 + (2/ \kappa)\,(1 + t) + ( 1 - 4 a)/ \kappa^2}}  \,-\,\frac{1}{\sqrt{(1 - t)^2}}\Big] \,\,-\,\,\int^{t^0}_0 d t  \frac{ t^{\ga - 1} - 1}{\sqrt{( 1 - t)^2}}\nn\\
 &+&\int^1_0 d t \left( t^{-\ga} - 1\right)\Big[\frac{1}{\sqrt{(1 - t)^2\,+\,(2/ \kappa)\,t (1 + t) \,+\,( 1 - 4 a)\,t^2/ \kappa^2}}\,-\,\frac{1}{\sqrt{(1 - t)^2}}\Big].
\end{eqnarray}
Then Eq.~(\ref{SC2}) can be cast into the form
\begin{equation} \label{SC8}
E\,=\,-\chi\left( \gamma\left( l\right)\right)\,\,+\,\,\tilde{T}\left( l, a\right) \,\,+\,\,CT\left( l,\gamma\left( l \right),a\right)\,\,- \,\,P\left( l, \gamma\left( l \right), a\right)  \,=\,- \chi\left( \gamma\left( l\right)\right)\,\,- \,\,\widetilde{P}\left( l, \gamma\left( l \right), a\right), \end{equation}
where
\begin{eqnarray} \label{CT}
CT\left( l, \gamma, a\right)\,\,\,&=&\,\,\frac{5}{9} \frac{e^{( 1 - \gamma)\,l}}{e^l\,+1\, - a}\left( a - 1\right)^{1 + \gamma} B\left( \frac{a - 1}{a}, 1 - \gamma,a\right),\\
\tilde{T}\left( l, a\right) \,\,&=&\,\,T\left( e^l - a\right) \,\,-\,\,L\left( e^l, a\right)\label{SC9}\\
\label{L}
L\left(\kappa, a\right)\,\,&=&\,\,\int^1_{\frac{a}{\kappa}} \frac{d t}{
  \sqrt{ (1 - t)^2 \,+\,\frac{2}{\kappa}( 1 + t) \,\,+\,\,\frac{1 - 4 a}{\kappa^2}}}
  \,\,+\,\, \int^1_{0} \frac{d t}{
  \sqrt{ (1 - t)^2 \,+\,\frac{2}{\kappa} t ( 1 + t) \,\,+\,\,( 1 - 4 a)\frac{t^2}{\kappa^2}}} \nn\\
  &=&-\ln\left( 1 + a -\kappa + \sqrt{(1 - a +\kappa)^2}\right) + 
\ln\left( 1 + \sqrt{   1 - 4 a + 
    4\kappa}\right)   - \frac{\kappa \ln \left(\kappa (1 - \kappa+ 
    \sqrt{-4 a + (1 +\kappa)^2}\right)}{
   \sqrt{-4 a + (1 +\kappa)^2}} \nn\\
&+&\frac{ \kappa \ln \left( 
   1 - 4 a + 3\kappa + 
 \sqrt{1 - 4 a + 4\kappa}
      \sqrt{-4 a + (1 + \kappa)^2}\right)}{\sqrt{-4 a + (1 + \kappa)^2}}
\end{eqnarray}
and $B\left( x, p, q\right)$ is the incomplete Beta function\footnote{See Eq.~{\bf 8.39} in Ref.~\cite{RY}}. Fig.~\protect\ref{ptct}  illustrates how all the ingredients of Eq.~(\ref{SC8}) behave as  functions of $l = \ln(\kappa + a(\nu))$.
      
      \section{Semi-classical solutions  to the BFKL equation with massive gluon: numerical results}
      \subsection{Trajectories and intercepts}
\label{subsec:TaI}
\begin{figure}[ht]
     \begin{tabular}{c c}
     \includegraphics[width=9.5cm]{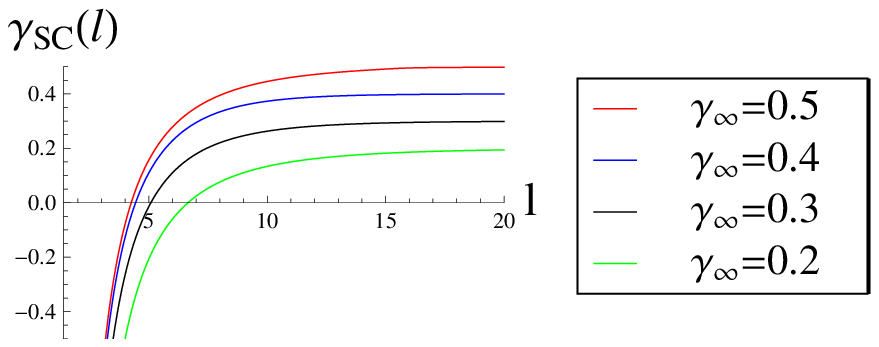} & \includegraphics[width=8cm]{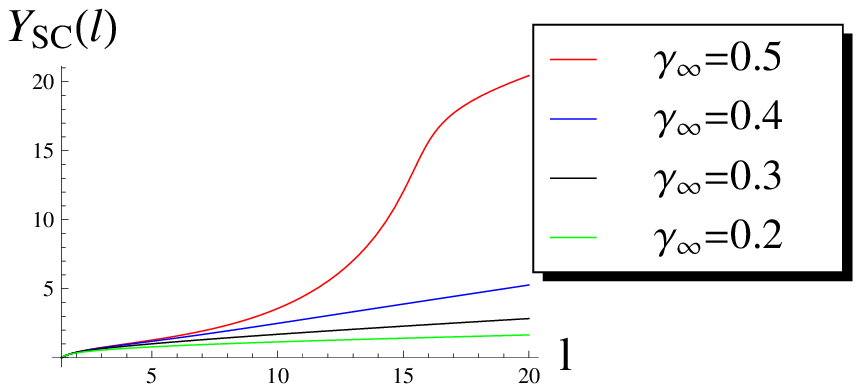} \\
      Fig.~\protect\ref{trajgen}-a &  Fig.~\protect\ref{trajgen}-b\\
         \end{tabular}
      \caption{ The trajectories for the BFKL equation with massive gluons (solutions to  Eq.~(\protect\ref{MC5})  and to  Eq.~(\protect\ref{MC7}) ) versus $l \,=\,\ln\left( \kappa + a\right)$. Function $Y_{SC}\left( l\right)$ is the inverse function to $l = l_{SC}\left( Y\right)$ of  Eq.~(\protect\ref{MC71}). }
\label{trajgen}
   \end{figure}

For solving the general BFKL equation with massive gluons, we need to find the trajectory from Eq.~(\ref{SP}) which will be a function of both $\xi = l/Y$ and $l$.  
According to the  method of characteristics discussed in Section~\ref{subsec:MC}, these trajectories are the solutions of Eq.~(\ref{MC5}) or, equivalently,  Eq.~(\ref{MC6}).
Unfortunately, we are able to solve Eq.~(\ref{MC5}) only numerically, and these solutions are shown in  Fig.~\protect\ref{trajgen}.

All the trajectories can be characterized by  their  asymptotic boundary condition $\gamma_{SC}\left( l \right)\,\,\xrightarrow{l\,\gg\,1} 
\gamma_\infty$. We  note that in Eq.~(\ref{CORE5}) one   should  understand $\gamma$ as $\gamma_\infty$, thus reducing it to the form
\begin{equation} \label{NR2}
T\left( - a\right)\,\,=\,\,-\,\,\chi\left( \gamma_\infty\right).
\end{equation}

\begin{figure}[ht]
     \begin{center}
     \includegraphics[width=13cm]{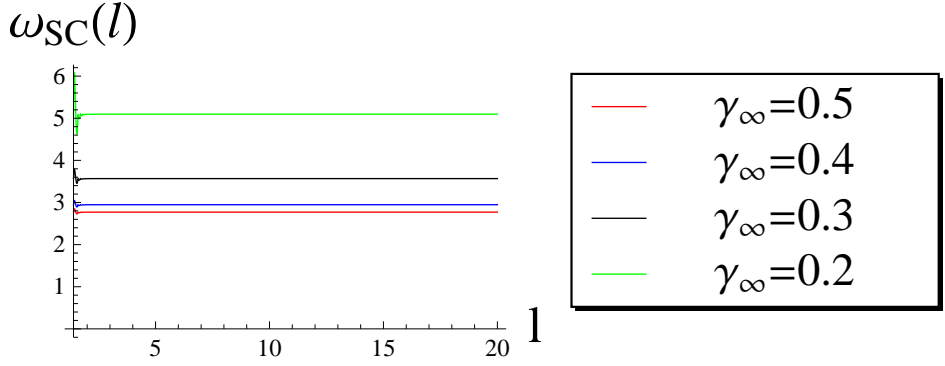} 
     \end{center}     
      \caption{ Intercept of the massive BFKL Pomeron versus $l$  on the trajectories with different $\gamma_\infty$.}
\label{omxi}
   \end{figure}

At small values of $l$ all the trajectories $\gamma_{SC}\left( l, \xi\right)$ vanish, reflecting the $K_0\left( \sqrt{a} r\right)$-behavior of the solution in the coordinate representation. The negative value of $\gamma_{SC} $ for soft $l$ in a numerical solution of Eq.~(\ref{MC5}) is different from what one expects from a massless BFKL. In section 3.4.3 we will discuss this region in more detail, here we only point out that $E\left( \gamma, l\right)$ is an analytical function of $\gamma$ for negative values of $\gamma$'s without any singularities at $\gamma = - n, n=0,1,2 \dots$. Fig.~\protect\ref{omxi} shows that $\gamma_{SP}\left( l \right)$ given by the solution of Eq.~(\ref{MC5}) and  presented in  Fig.~\protect\ref{trajgen}, satisfy  the equation even at $l \,<\,l_{soft}$.
 
  The trajectories $l_{SP}\left( Y\right)$ can be found by solving Eq.~(\ref{MC71}), and are shown in Fig.~\protect\ref{trajgen}-b for several boundary conditions. In the $\xi \ll 1$ region the solution may be constructed analytically without solving an additional equation.  The first observation is that at large $l$, the set of trajectories should coincide with the same set for the massless BFKL equations~(\ref{SCM0}),
 \begin{equation} \label{NR3a} 
   \gamma_{SP}\left( l, \xi\right) \,\xrightarrow{l \,\gg\,1}\,\gamma^{\mbox{\tiny BFKL}}_{SP}\left( \xi\right)\equiv \gamma_\infty.
   \end{equation}
   
 Assuming that $\gamma^{\mbox{\tiny BFKL}}_{SP}\left( \xi\right)$ is close to $\frac{1}{2}$ at small $\xi$ (in so called Bjorken limit), we can consider a deviation as a small parameter,
   
\begin{equation} \label{NR1}
\gamma_{SP}\left( l \right)\,\,=\,\,\gamma_{SC}\left( l, \gamma_\infty=\frac{1}{2}\right)\,\,+\,\,\delta\gamma_{SP}\left( l, \xi\right).
\end{equation}
To find  $\delta\gamma_{SP}\left( l,\xi\right)$,  Eq.~(\ref{SP}) can be re-written in the form
\begin{equation} \label{NRSP}
\delta\gamma_{SP}\left( l\right)\,=\,
\lim_{\epsilon \to 0}\Bigg( - \xi\,\frac{\Big( \gamma_{SC}\left( l, \gamma_\infty=\frac{1}{2}\right)\,+\,\epsilon \,-\,\frac{1}{2}\Big)}{\Big( \frac{d \chi\left( \gamma = \gamma_{SC}\left( l, \gamma_\infty=\frac{1}{2}\right)\,+\,\epsilon\right)}{d \gamma}\,\,+\,\,\frac{\partial \widetilde{P}\left( l, \gamma=\gamma_{SC}\left( l, \gamma_\infty=\frac{1}{2}\right)+\,\epsilon, a \left( \gamma_\infty\right)\right)}{\partial \gamma}\Big)}\Bigg)
\end{equation}
where $\epsilon\approx 0$ is a small cutoff needed to regularize some intermediate results. In particular, Eq.~(\ref{NRSP}) reproduces Eq.~(\ref{DIFA1}) and gives the  final result at $l \,\gg \,1$ where $\frac{d \chi\left( \gamma = \gamma_{SP}\left( l \right)\right)}{d \gamma}\,\,+\,\,\frac{\partial \widetilde{P}\left( l, \gamma=\gamma_{SC}\left( l \right), a \left( \gamma_\infty\right)\right)}{\partial \gamma}\,\,\ll\,\,1$.

The solution to Eq.~(\ref{SP}) in this approximation has a  form
\begin{eqnarray} \label{NR3}
 \Psi\left( Y, l \right)\,\,&=&\, \phi_{in}\left( \ga_{SP}\left( Y, l\right)\right)\,\sqrt{\frac{1}{ 2 \pi \, \Big{|}\frac{\partial^2 E\left( \ga_{SP}\left( Y, l \right) , l\right)}{\partial \ga^2_{SP}}\Big{|}\, Y}}\,\,e^{-\frac{1}{2} l}\,\, e^{\omega_{eff}\left( l, \xi\right) Y},\\
 \end{eqnarray}
where we introduced an effective intercept
\begin{equation}
 \omega_{eff}\left( l, \xi\right) =  - E\left( \ga_{SP}\left( Y, l\right), l\right) + \left( \ga _{SP}\left( Y, l\right)- \frac{1}{2}\right) \xi,
\end{equation}
and $\gamma_{SP}$ is given by Eq.~(\ref{NR1}).

\begin{figure}[ht]
     \begin{tabular}{c c}
     \includegraphics[width=8cm]{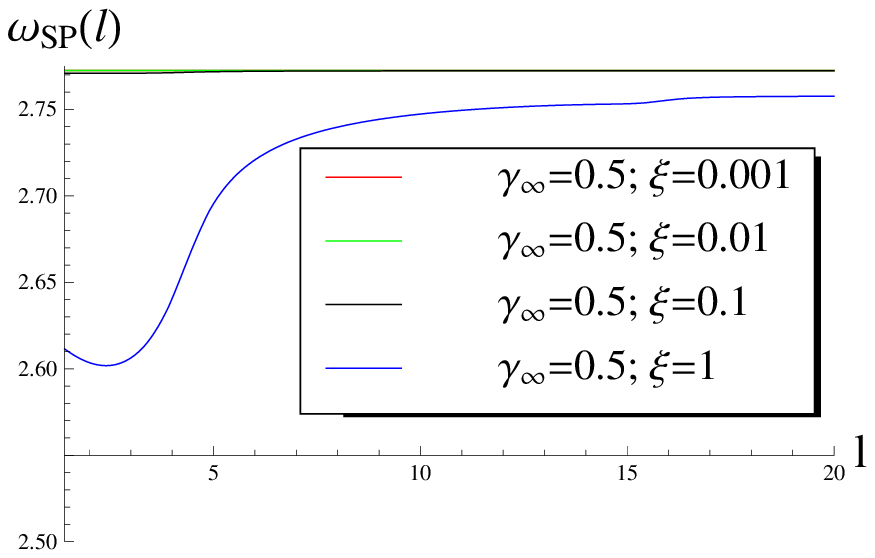} &    \includegraphics[width=8cm]{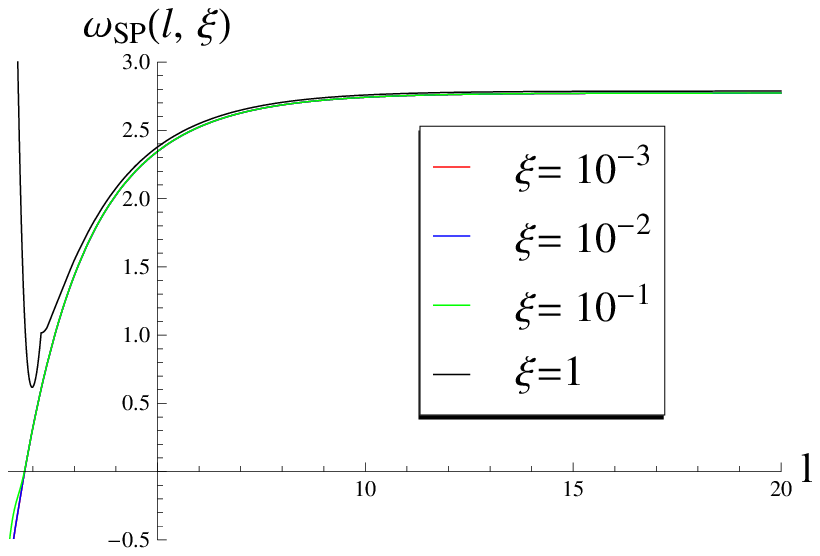}\\
      Fig.~\protect\ref{omsp}-a& Fig.~\protect\ref{omsp}-b\\
          \end{tabular}     
      \caption{ $\omega_{eff} \,\equiv\,\omega_{SP}$  of the massive BFKL Pomeron versus $l $ at fixed $\xi = l/Y$.\protect Fig.~\protect\ref{omsp}-b illustrates that $\omega_{SP}\left( l \right)$ are close for small values of $\xi$.}
\label{omsp}
   \end{figure}

 The $l$-dependence of the effective intercepts $\omega\left( l, \xi\right)$ is shown in  Fig.~\protect\ref{omsp}. Note  at small $\xi $,  these intercepts are close to the massless BFKL  given by  Fig.~\protect\ref{m0}-a in the entire region of $l$.  In the region of large  $\xi$, the effective intercepts are considerably smaller than the intercept of the massless BFKL Pomeron. For small $l$, \textit{i.e.}, the scattering amplitude at small values of $l$ and large values of $\xi$,  is suppressed in comparison to  the behavior determined by the intercept of the massless BFKL Pomeron. Since, the asymptotic region at high energies (large $Y$) corresponds to $\xi \,\ll\,1$, the high energy behavior of the massive BFKL Pomeron is the same as for the massless one. Therefore, we have  the confirmation of our basic results of the numerical calculation (see Ref.~\cite{LLS}) in the framework of the semiclassical approach. 

\subsection{Accuracy of the semiclassical approach}
\begin{figure}[ht]
     \begin{tabular}{c c c}
     \includegraphics[width=7.5cm]{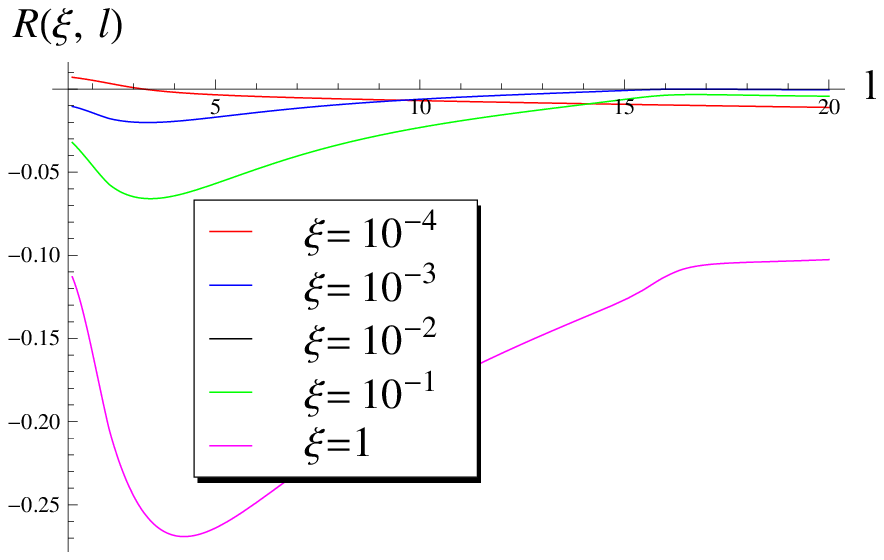} &~~~~~~~&  \includegraphics[width=7.5cm]{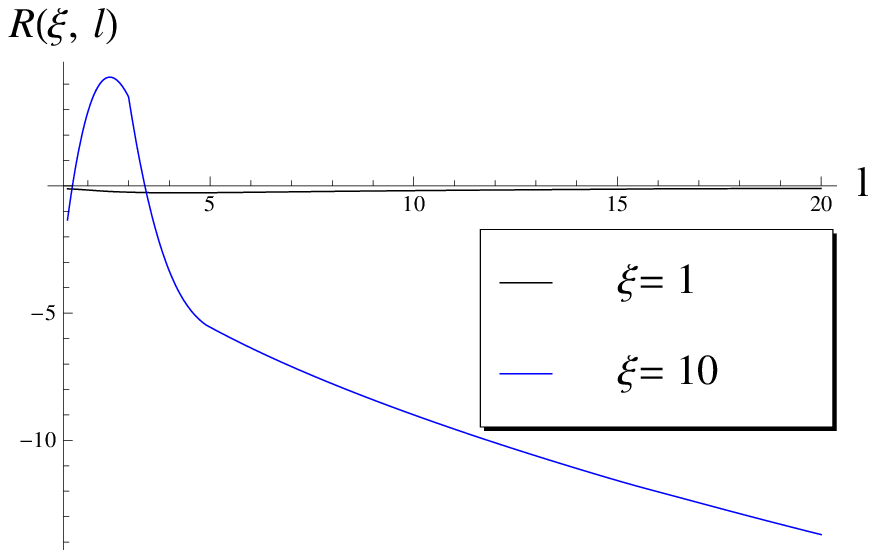} \\
      Fig.~\protect\ref{r3}-a &  & Fig.~\protect\ref{r3}-b \\
     \end{tabular}     
      \caption{ The ratio of  Eq.~(\protect\ref{SCCON}) for large values of $Y$. \protect Fig.~\protect\ref{r3}-a shows this ratio at fixed $\xi = l/Y$ which are small, and \protect Fig.~\protect\ref{r3}-b presents $R$ at $\xi =1$ as a function of $l$.}
\label{r3}
   \end{figure}

   The accuracy of the semiclassical approximation is controlled by the ratio $R$ of Eq.~(\ref{SCCON}).  Fig.~\protect\ref{r3} shows  this ratio for different values of $\xi$. From the figure we conclude that for $\xi \,\leq\,1$ the ratio $R$ is small and we can safely use the semiclassical approximation. 
 
 For $l < 4$, the ratio $R$ is small in the region of large $Y$ (small $\xi$). The reason for this is since for large $l$, the solution of the BFKL equation for the massive gluons,  should coincide with the solution of the massless BFKL equation.
These estimates confirm our expectation that the semiclassical method provides a reliable approach at high energies (large value of $Y$).

At large $\xi$,  our procedure does not work even at large $l$. In terms of  kinematics,  we need to deal with $\gamma_{SC}  \ll 1$, while in the Fig.~\protect\ref{r3} we used $\gamma_{SC} \,\xrightarrow{ l \,\gg\,1} \,\frac{1}{2}$. For this region of large $l$ and $\xi$,  we develop an approximation which corresponds to the double log approximation and for very large $l$ , and coincides with the DLA for the massless BFKL equation.

\begin{figure}[ht]
     \begin{center}
     \includegraphics[width=10.5cm]{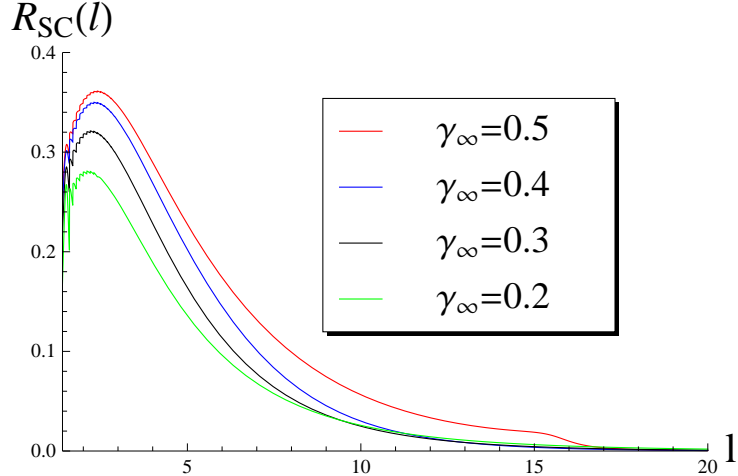}  
     \end{center}     
      \caption{ The ratio of  Eq.~(\protect\ref{RSC})  versus $l$. \.}
\label{rsc}
   \end{figure}

In Fig.~\protect\ref{rsc} we plot the ratio
\begin{equation} \label{RSC}
R_{SC}\,\,=\,\, \frac{\frac{d \gamma_{SC}\left( l \right)}{d l} }{\left(  1 \,-\, \gamma_{SC}\left( l \right)\right)^2}.
\end{equation}
This parameter controls the smoothness of the functions $\omega\left( Y,l\right)$ and $\gamma\left( Y, l\right)$, and thus a precision of the semiclassical approach. The Fig.~\protect\ref{rsc} shows that this ratio is very small at large $l$, and doesn't exceed $ \approx 0.4$ even at small $l$.
\subsection{Saturation momentum}
\label{subsec:SatMom}
\begin{figure}[ht]
     \begin{tabular}{c c}
     \includegraphics[width=9cm]{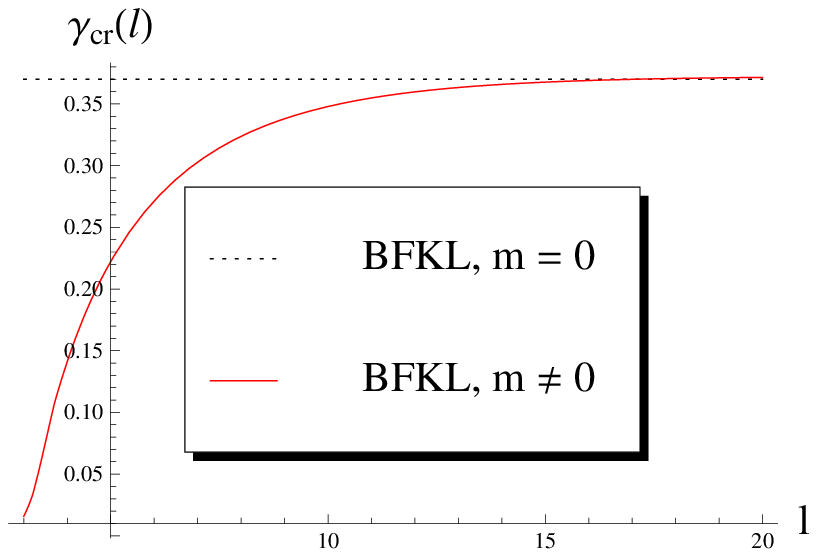} &  \includegraphics[width=7.5cm]{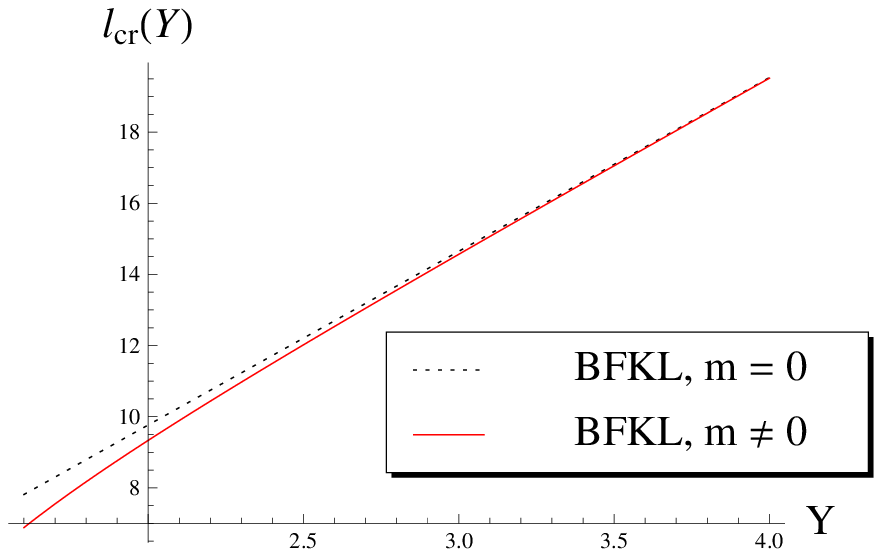} \\
      Fig.~\protect\ref{qs}-a & Fig.~\protect\ref{qs}-b \\
     \end{tabular}     
      \caption{ The critical trajectories for BFKL equation with $m =0$ and with $m \neq 0$ (see \protect Fig.~\protect\ref{qs}-a).  \protect Fig.~\protect\ref{qs}-b shows the dependence of the saturation scale ($\ln\left( Q^2_s\left( Y \right)/ Q^2_s\left( Y = 0 \right)\right)$) as function of $Y$. }
\label{qs}
   \end{figure}

It is well known that we can find a saturation momentum by  searching for the particular trajectory on which the wave function $\Psi\left( Y, \kappa\right) = \mbox{const}$ (see Refs.~\cite{GLR,MUTR,MUPE}). Mathematically, this corresponds to a solution of a system of two equations
  \begin{eqnarray} \label{QS1}
\mbox{trajectory:} &&   - \frac{ \partial E\left( \ga_{SP}\left( Y, l \right) , l\right) }{\partial \ga}\,Y\,+\, l \,=\,0,\nn\\
\mbox{front line:}  &&  E\left( \ga_{SP}\left( Y, l \right) , l\right)\,Y  \,=\,(1 - \gamma_{SP})\,l.
\end{eqnarray}
  In the case of massless the BFKL equation, the solution to the equations of Eq.~(\ref{QS1}) is $\gamma_{SP} = \gamma_{cr} \,\,=\,\,0.37$ ~\cite{GLR}. For the massive BFKL equation, the critical trajectory is shown in  Fig.~\protect\ref{qs}-a. The equation for the saturation momentum for massless BFKL equations of the form
  \begin{equation} \label{QS2}
  l_{cr}\,\,=\,\,\ln\left( Q^2_s\left( Y \right)/ Q^2_s\left( Y = 0 \right) \right) \,\,=\,\,\frac{\chi\left( \gamma_{cr}\right)}{1 - \gamma_{cr}}\,Y.
  \end{equation}
  The solution to Eq.~(\ref{QS1}) for $l_{cr} = \ln\left( Q^2_s\left( Y \right)/ Q^2_s\left( Y = 0 \right)\right)   $ is shown in  Fig.~\protect\ref{qs}-b.
The difference between these two cases is sizable,  only for small values of $l = \ln\left( \kappa + a\right)$.

 \subsection{Analytical solutions}
In this section we develop two analytical methods of searching for  solutions based on the diffusion and DLA approximations ,for the massless BFKL equation.
\subsubsection{Diffusion approximation}
\begin{figure}[ht]
     \begin{tabular}{c c}
     \includegraphics[width=8.7cm]{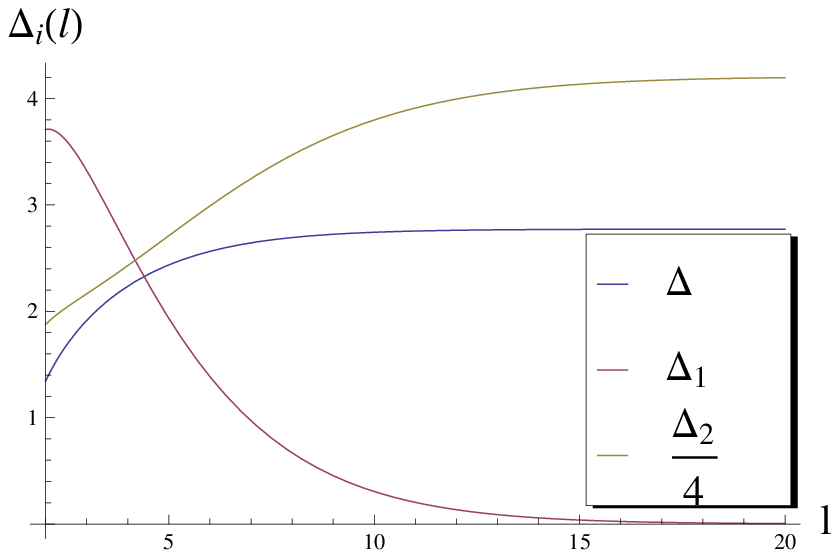} &  \includegraphics[width=8.7cm]{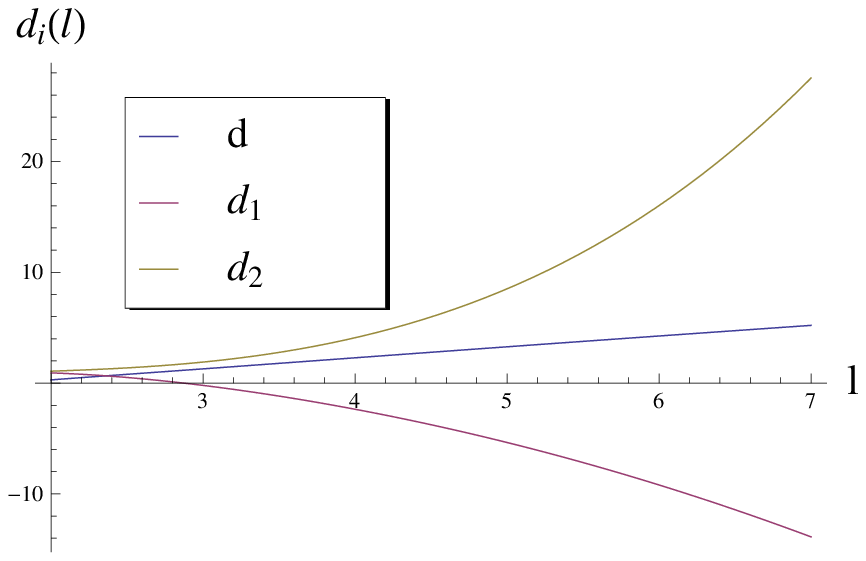}\\
      Fig.~\protect\ref{deltas}-a & Fig.~\protect\ref{deltas}-b\\
          \end{tabular}    
      \caption{ Functions $\Delta_i\left( l \right)$ of  Eq.~(\protect\ref{DAGC}) (see \protect Fig.~\protect\ref{deltas}-a) and functions $d_i\left( l \right)$ in  Eq.~(\protect\ref{LGA4}) ( see \protect Fig.~\protect\ref{deltas}-b)}
\label{deltas}
   \end{figure}

\begin{figure}[ht]
     \begin{center}
     \includegraphics[width=9cm]{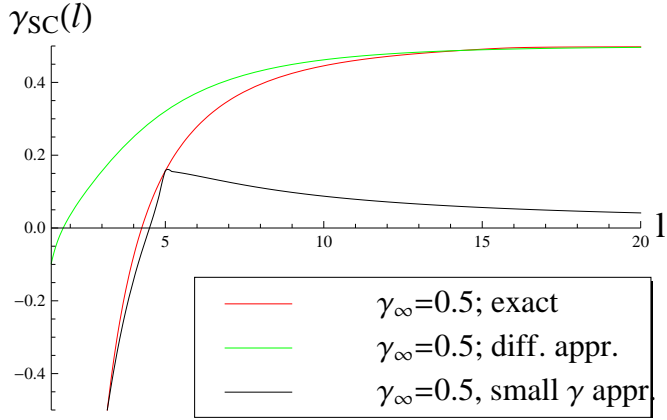} 
     \end{center}    
      \caption{ Functions $\gamma_{SC}\left( l, \xi \right)$  versus $l$: exact solution to  Eq.~(\protect\ref{MC5}); diffusion approximation and approximation at small values of $\gamma$. In the figure $\xi =0.01$.}
\label{gaan}
   \end{figure}

A brief glance at the trajectories  for small values of $\xi$ (see  Fig.~\protect\ref{omxi}) allows us to conclude that these trajectories are close to $\ga = \frac{1}{2}$ at least for $l \geq 5$. Therefore, for such values of $l$ we can develop the diffusion approach, in complete analogy with the case of massless BFKL equation, that has been discussed in section~\ref{sub:DiffApprox}. In the vicinity of $\ga = \frac{1}{2}$ we can expand the general expression of Eq.~(\ref{SC8}) as
\begin{equation} \label{DAGC}
\omega\left( \gamma, l\right)\,\,=\,\,- E\left( \gamma, l,a\right)\,\,=\,\,\Delta\left( l \right)\,\,+\,\,\Delta_1\left( l \right) \left( \gamma - \frac{1}{2}\right)\,\,+\,\,\Delta_2\left( l \right) \left( \gamma - \frac{1}{2}\right)^2.
\end{equation}
Functions $\Delta\left( l \right), \Delta_1\left( l \right)$ and $\Delta_2\left( l \right)$ are plotted in  Fig.~\protect\ref{deltas}-a.
 We see that at large $l$ (say at $l \geq  l_0 \,\approx \,5$) the functions $\Delta_i$ reach constant values, $ \Delta \left( l \right) \xrightarrow{l > l_0} \, \omega_0$; $ \Delta_1 \left( l \right) \xrightarrow{l > l_0} \, 0$;  and  $ \Delta_2 \left( l \right) \xrightarrow{l > l_0} \, D$.
 Substituting expansion~(\ref{DAGC}) into Eq.~(\ref{MC5}), we obtain
 \begin{equation} \label{DASC1} 
 \Delta \left( l \right) \,+\,\Delta_1\left( l \right) \,\left( \gamma_{SC}\left( l \right)\,-\,\frac{1}{2}\right)  \,\,+\,\,\Delta_2\left( l \right) \left( \gamma_{SP} - \frac{1}{2}\right)^2\,\,=\,\,\chi\left( \frac{1}{2} \right),
 \end{equation}
 whose solution is
 \begin{equation} \label{DASC2}
 \gamma_{SC}\left( l \right)\,\,\equiv \,\,\gamma_D\left( l \right)\,\,=\,\,\left( - \Delta_1\left( l \right)\,\pm\,
 \sqrt{ \Delta^2_1\left( l \right)\,-\,4\,\Delta_2\left( l \right) \left(  \Delta\left( l \right) \,-\,\chi\left( \frac{1}{2}\right)\right)}\right){\Big/}\Big(2 \,\Delta_2\left( l \right) \Big).
 \end{equation} 
The  physical $\gamma_D$ corresponds to the  minus  sign in Eq.~(\ref{DASC2}). In  Fig.~\protect\ref{gaan}  we compare the trajectory Eq.~(\ref{DASC2}) with the exact trajectory at $\xi = 0.01$,  that has been calculated in Section~\ref{subsec:TaI}. For $l > l_0 \sim 10$ we can safely use the solution Eq.~(\ref{DASC2}).
 
 Using Eq.~(\ref{DASC1}), we can calculate the wave function of Eq.~(\ref{SP1}) which takes the form
 \begin{equation} \label{DAWF}
 \Psi\left( Y, l \right)\,\,=\,\,\phi_{in}\left(  \gamma_{SP}\left( l \right)\right) \,\sqrt{\frac{1}{ 4 \it \Delta_2\left( l \right) Y}}\,e^{ - \frac{1}{2} l}\, e^{\omega_{eff}\left( l, \xi\right)\,Y},
 \end{equation}
 with
 \begin{equation} \label{OMDA}
  \omega_{eff}\left( l, \xi\right)\,\,\,=\,\,\,\chi\left( \frac{1}{2} \right) \,\,-\,\,\frac{\left( \xi \left( \gamma_D\left( l \right) - \frac{1}{2}\right)\right)^2}{2\left( \Delta_1\left( l \right) + 2 \, \Delta_2\left(  l \right)\,\left( \gamma_D\left( l \right) - \frac{1}{2} \right)\right)}.
  \end{equation}
  The Eq.~(\ref{OMDA}) provides a good description of the intercept for rather large $l \,>\,l_0 \sim 10$.

\subsubsection{Small $\gamma$ approximation}
For $l \,<\,l_0$,  we cannot use the diffusion approximation since, as we can see from Fig.~\protect\ref{gaan}, the diffusion trajectory is much larger  than the exact one in this region.  Actually, the exact $\gamma_{SP}\left( l, \xi\right)$ at small $\xi$ approaches $\gamma_{SP} \to 0$. At $\gamma \to 0$ the general expression for $P\left(l, \gamma,a\right)$ in Eq.~(\ref{SC8}) can be simplified and takes the form
\begin{eqnarray} \label{LGA1}
&&\chi\left( \gamma\right) + \tilde{P}\left( l,\gamma, a\right)\,\,\xrightarrow{\gamma \to 0}\,\,\frac{1}{\gamma} + \tilde{P}_{\gamma \to 0}\left( l,\gamma,a\right)\,\,=\,\,\frac{1}{H\left( l, a\right)}\frac{1}{\gamma}\left( 1 - e^{- \gamma\left( l - l_a\right)}\right),\\
&& \mbox{with} ~~~H\left( l, a\right) =\,1/\sqrt{\left( 1\, + \,\exp\left( - l \right)\right)^2   - 4 a \exp\left(- 2 l \right)}\,\,\,\,\mbox{and}~~~l_a = \ln a, \nn
\end{eqnarray}
The Eq.~(\ref{SP}) takes the form
\begin{equation} \label{LGA2}
\frac{1}{H\left( l, a\right)}\Big\{ - \frac{1}{\gamma_{SC}^2}\left( 1 - e^{- \gamma_{SC}\left( l - l_a\right)}\right) \,+\,\left( l - l_a\right)
\frac{1}{\gamma_{SC}} \,e^{- \gamma_{SC}\left( l - l_a\right)}\Big\}\,\,=\,\,-\,\,\xi.
\end{equation}

The Eq.~(\ref{LGA2}) has two  solutions in different regions: (1) $\gamma\,\ll\,1$ but $\gamma\left( l - l_a\right) \,\gg\,1$; and (2)  $\gamma\,\ll\,1$ and  $\gamma\left( l - l_a\right) \,\ll\,1$.

 In the first kinematic region Eq.~(\ref{LGA2}) reduces to
\begin{equation} \label{LGA3}
 - \frac{1}{\gamma_{SC}^2}\,\,=\,\,H\left( l, a\right),
 \end{equation}
with the solution
\begin{equation}
 \gamma_{SC}\,\,=\,\,\sqrt{H\left( l, a \right) \,\xi}.
\end{equation}
 We can check that 
$\gamma\left( l - l_a\right)  \,\,=\,\,\frac{1}{\sqrt{H\left( l, a \right) \,\xi}}\left( l - l_a\right) \xrightarrow{l \gg l_a}\,\sqrt{\frac{ Y\, l}{H\left( l, a \right)}}\,\,\gg\,\,1.$
Therefore, this solution corresponds to the DLA approximation in this kinematic region.
 
 For the kinematic region $\gamma\,\ll\,1$ and  $\gamma\left( l - l_a\right) \,\ll\,1$  the Eq.~(\ref{LGA1}) leads to the analytical function at $\gamma \to 0$ in contrast to the case of the massless BFKL kernel.  Therefore, we  can search for  a  parametrization  $\omega\left( l, \xi\right)$,  which has    the same form as Eq.~(\ref{DASC1}), \textit{viz.:}
 \begin{equation} \label{LGA4}
\omega_{\gamma\, \ll\,1}\left( \gamma, l\right)\,\,=\,\,- E\left( \gamma, l,a = 4\right)\,\,=\,\,d \left( l \right)\,\,+\,\,d_1\left( l \right)  \gamma\,\,+\,\,d_2\left( l \right) \, \gamma^2,
\end{equation} 
  with functions $d_i$ is plotted in  Fig.~\protect\ref{deltas}-b. We can see from  Fig.~\protect\ref{gaan} that we can rely on Eq.~(\ref{LGA4}) only for $l - l_a \ll 1/(\gamma = 0.2) \approx 5$.
 
 Eq.~(\ref{SP}) with $\omega\left( l, \xi\right)$ given by Eq.~(\ref{LGA4}),  has a solution
 \begin{equation} \label{LGA5}
 \gamma_{SC}\left( l \right)\,\,\equiv \gamma_S\left( l \right)\,\,\equiv \,\,\gamma_D\left( l \right)\,\,=\,\,\left( - d_1\left( l \right)\,\pm\,
 \sqrt{ d^2_1\left( l \right)\,-\,4\,d_2\left( l \right) \left(  d\left( l \right) \,-\,\chi\left( \frac{1}{2}\right)\right)}\right){\Big/}\Big(2 \,d_2\left( l \right) \Big).
 \end{equation}  
    The solution~(\ref{LGA5}) is shown in  Fig.~\protect\ref{gaan}. This solution leads to good approximation for $l  = 3 \div 5$ with
    \begin{equation} \label{LGA6}
  \omega_{eff}\left( l, \xi\right)\,\,\,\approx\,\,\,\chi\left( \frac{1}{2} \right) \,\,-\,\,\frac{\xi^2 \, \gamma^2_S\left( l \right) }{2\left( d_1\left( l \right) + 2 \, d_2\left(  l \right)\, \gamma_S\left( l \right) \right)}.    \end{equation}
    The wave function takes the form
     \begin{equation} \label{LGA7}
 \Psi\left( Y, l \right)\,\,=\,\,\phi_{in}\left(  \gamma_{SP}\left( l \right)\right) \,\sqrt{\frac{1}{ 4 \,d_2\left( l \right) Y}}\,e^{ -  l}\,e^{ \omega_{eff}\left( l \right)\,Y}.
 \end{equation}

\begin{figure}[ht]
     \begin{center}
     \includegraphics[width=9cm]{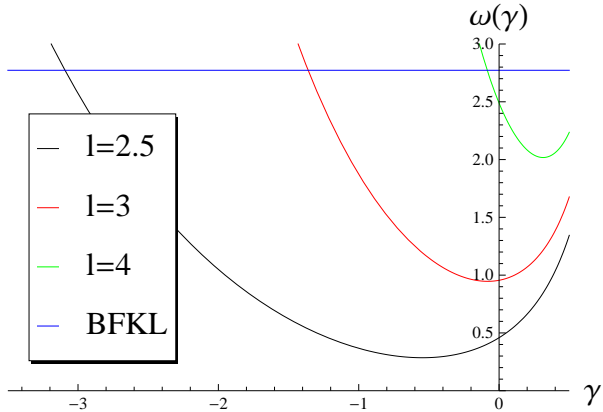} 
     \end{center}    
      \caption{\label{omga}Function $\omega\left( \gamma, l\right) $ versus $\gamma$ at fixed $l$. $\omega_{BFKL}\,\,=\,\,\chi\left( \frac{1}{2} \right)$.}
   \end{figure}

 {\boldmath
\subsubsection{  $l\,\leq \,l_{soft}$ }}
    Earlier we have seen that the trajectory $\gamma_{SP}(l)$ has a node at some value $l_{soft}$,  and becomes even negative for $l\lesssim l_{soft}$.  As we have seen earlier, the function $\omega_{SP}(\gamma)$ is analytic at $\gamma=0$, so we can extrapolate  to negative but small $\gamma$'s using Eq.~(\ref{LGA4}) (see  Fig.~\protect\ref{omga}).
  In principle, we can expect some high twists singularities at $\gamma = - n $ with $ n = 0, 1, 2,\dots$ which stem from the expansion of
  \begin{equation} \label{SOFT1}
  \frac{1}{\sqrt{( 1 - t)^2 + (2/ \widetilde{\kappa})\,(1 + t) + ( 1 - 4 a)/ \widetilde{\kappa}^2}} \,\,=\,\,\sum_{n = 0}^{\infty}\,C_n\left( l, a\right) \,t^n 
  \end{equation}
  at small $t$ in Eq.~(\ref{SC7}).    Taking the integral over $t$ in Eq.~(\ref{SC7}) in the vicinity of small $t$,  one can see that for $\gamma \to -n$  we obtain the same expression as in Eq.~(\ref{LGA1}),  replacing $\gamma$ in Eq.~(\ref{LGA1}) by $\gamma + n$. The energy (intercept) turns out to be a regular function at $\gamma = -n$, and we can use Eq.~(\ref{LGA4})  for $| \gamma + 1|\,<\,1$
 with function $d^{(1)}_i\left( l \right)$ calculated at $\gamma = -1$ ( see  Fig.~\protect\ref{gaan} and  Fig.~\protect\ref{omga}).  
    
      Fig.~\protect\ref{omga} illustrates that  the intercept $\omega\left( \gamma,l\right)$ is the analytical function without singularities at negative $\gamma$ which increases at large $|\gamma|$.   

 Unfortunately, we have not found a simple analytical approach that is able to describe the scattering amplitude at all values of $l$. However, we would like to recall that the numerical solutions to Eq.~(\ref{NR1}) and Eq.~(\ref{NRSP}) depend neither on the value of the QCD coupling, nor on the initial condition, and reduce the  procedure of calculation of the scattering amplitude to a simple equation. Solving this equation is much easier task than the exact numerical calculation of the eigenvalues and eigenfunctions that was done in Ref.~\cite{LLS}.

 \section{Conclusions}   
   In our previous paper (see Ref.~\cite{LLS}) we studied the BFKL equation with massive gluons in the lattice and proved that its spectrum coincides with the spectrum of the massless BFKL equation. This observation gives rise to a hope that the correct large impact parameter ($b$) behavior of the scattering amplitude $A \propto  \exp\left( - m b\right) $, which is the inherent feature of the massive BFKL equation, will not affect the high energy behavior of the scattering amplitude. Therefore, we can expect that the modification of the BFKL equation due to confinement  would  not affect strongly the main equations that govern the physics at high energy ( in particular, the non-linear equations of the  CGC/saturation approach to high density QCD).    
   
   In this paper we developed the semiclassical approximation which allowed us to investigate the high energy behavior of the scattering amplitude. The method provides a simple procedure for the calculation and reduces it to a numerical solution of Eq.~(\ref{MC5}), which is much simpler than the direct numerical calculations of  the eigenvalue problem  in the lattice realized in Ref.~\cite{LLS}.
   
    Having these solutions, we  propose a modification of  high energy asymptotic behavior, caused by the correct large $b$ exponential fall off of the amplitude.  Actually, we did not find any unexpected behavior, and the semiclassical solution reproduces the scattering amplitude which is very close to the amplitude of the massless BFKL equation, at least at high energies.
    
   In section~\ref{subsec:SatMom} we estimate the value of the saturation momentum, solving the linear evolution equation with very general assumptions about the non-linear corrections. We demonstrated that the value of the saturation momentum is close to the one for the massless BFKL equations, leading to the assumption that  saturation physics will look similar for both massive and massless BFKL equations.
   
 We believe that in this paper we taken the natural next step in the understanding of the influence of the correct large $b$ decrease of the amplitude on its high energy behavior. It should be stressed that this behavior  is interesting as it provides the high energy amplitude for the electroweak-weak theory, which can be measured experimentally.   The solution which has been discussed in this paper, determines the asymptotic high energy behavior of the electroweak-weak theory for zero Weinberg angle.  
We plan to address the physical case of nonzero Weinberg angle~\cite{BLP} elsewhere.

 \section{Acknowledgements}
  
 We thank our    colleagues at UTFSM and Tel Aviv university for encouraging discussions.   This research was supported by the BSF grant 2012124  and by the  Fondecyt (Chile) grants  1140842 and 1140377.

    \end{document}